\documentclass[sigconf]{acmart}

\renewcommand\footnotetextcopyrightpermission[1]{} 

\setcopyright{none}

\usepackage{color}
\usepackage{url}
\usepackage{verbatim} 
\usepackage[FIGTOPCAP]{subfigure} 
\usepackage{multirow}
\usepackage{algorithm}
\usepackage[noend]{algorithmic}
\usepackage{xspace}
\usepackage{graphicx}
\usepackage{epsfig}
\usepackage{amsmath}
\usepackage{amssymb}
\usepackage{float}
\usepackage{enumitem}
\usepackage{balance}
\usepackage[font=bf,belowskip=0pt,aboveskip=5pt]{caption}
\usepackage{lipsum}
\usepackage{bm}
\usepackage{hyperref}
\hypersetup{colorlinks,citecolor=blue}

\usepackage{amsmath}
\makeatletter
\g@addto@macro\normalsize{%
  \setlength\abovedisplayskip{2pt}
  \setlength\belowdisplayskip{2pt}
  \setlength\abovedisplayshortskip{2pt}
  \setlength\belowdisplayshortskip{2pt}
}

\usepackage{epigraph} 

\usepackage{etoolbox}
\usepackage{booktabs} 

\newcommand{\hide}[1]{}
\newcommand{\xhdr}[1]{\vspace{1.0mm}\noindent{{\bf #1.}}}

\newcommand{\eg}{\emph{e.g.}}
\newcommand{\ie}{\emph{i.e.}}

\DeclareMathOperator*{\argmax}{argmax}

\setitemize{noitemsep,topsep=0pt,parsep=0pt,partopsep=0pt,leftmargin=*}

\graphicspath{{./FIG/}}

\begin{document}

\copyrightyear{2018}
\acmYear{2018}
\setcopyright{iw3c2w3}
\acmConference[WWW 2018]{The 2018 Web Conference}{April 23--27, 2018}{Lyon, France}
\acmBooktitle{WWW 2018: The 2018 Web Conference, April 23--27, 2018, Lyon, France}
\acmPrice{}
\acmDOI{10.1145/3178876.3186161}
\acmISBN{978-1-4503-5639-8/18/04}

\title{Modeling Interdependent and Periodic\\Real-World Action Sequences}
\renewcommand{\shorttitle}{Modeling Interdependent and Periodic Real-world Action Sequences}

\author{Takeshi Kurashima}
\affiliation{%
\institution{NTT Corp. \& Stanford University}
}
\email{kurashima.takeshi@lab.ntt.co.jp}

\author{Tim Althoff}
\affiliation{%
\institution{Stanford University}
}
\email{althoff@cs.stanford.edu}

\author{Jure Leskovec}
\affiliation{%
\institution{Stanford University}
}
\email{jure@cs.stanford.edu}

\begin{abstract}

Mobile health applications, including those that track activities such as exercise, sleep, and diet, are becoming widely used. 
Accurately predicting human actions in the real world is essential for targeted recommendations that could improve our health and for personalization of these applications.
However, making such predictions is extremely difficult due to the complexities of human behavior, which consists of a large number of potential actions that vary over time, depend on each other, and are periodic. Previous work has not jointly modeled these dynamics and has largely focused on item consumption patterns instead of broader types of behaviors such as eating, commuting or exercising. 

In this work, we develop a novel statistical model, called \emph{TIPAS}, for Time-varying, Interdependent, and Periodic Action Sequences. Our approach is based on personalized, multivariate temporal point processes that model time-varying action propensities through a mixture of Gaussian intensities. 
Our model captures short-term and long-term periodic interdependencies between actions through Hawkes process-based self-excitations. 
We evaluate our approach on two activity logging datasets comprising 12 million real-world actions (\eg, eating, sleep, and exercise) 
taken by 20 thousand users over 17 months. We demonstrate that our approach allows us to make successful predictions of future user actions and their timing. Specifically, TIPAS improves predictions of actions, and their timing, over existing methods across multiple datasets by up to 156\%, and up to 37\%, respectively. Performance improvements are particularly large for relatively rare and periodic actions such as walking and biking, improving over baselines by up to 256\%. This demonstrates that explicit modeling of dependencies and periodicities in real-world behavior enables successful predictions of future actions, with implications for modeling human behavior, app personalization, and targeting of health interventions.

\end{abstract}

\maketitle

\section{Introduction}
\label{sec:intro}

\enlargethispage{\baselineskip}

Activity tracking applications for mobile health have become an important part of people's daily lives.
A US-nationwide study in 2013 found that 69\% of adults keep track of a health indicator, 
and 21\% among them used an app or device to do so~\cite{fox2013tracking}.
In activity logging applications such as Fitbit, Under Armour Record, and Argus, users might take one of many possible actions from a large and diverse space of potential actions at any point in time. 
Users continuously track many actions of their lives including exercise, diet, sleep, and commuting behavior with the goal of improving self-knowledge and personal well-being~\cite{swan2013quantified,althoff2017onlineactions,althoff2017large,shameli2017gamification}. 
User modeling is critical to making activity logging applications more useful by providing users with personalized experiences matching their specific objectives~\cite{fischer2001user,zukerman2001predictive,gorniak2000predicting,du2016recurrent,berkovsky2008mediation}. 
This has the potential to significantly improve people's health, 
for instance by preventing negative health outcomes and promoting the adoption and maintenance of healthy behaviors~\cite{nahum2016just,thomas2015behavioral,freyne2010intelligent,althoff2017population}.
However, successful personalization of systems rests on the ability to predict the user's next actions and when they will occur~\cite{zukerman2001predictive,davison1998predicting,du2015time}. 

Predicting actions is important because these predictions facilitate personalization of the user interface and user experience in order to provide users with what they need, without them asking for it explicitly~\cite{mulvenna2000personalization}. 
For example, in activity logging applications we can predict when the user will eat dinner and their future location in order to provide relevant recommendations~\cite{yu2016geographic}. 
Accurate and contextualized predictions could further help users to realize their personal goals by reminding them to measure their weight or notifying them about the exercise the next morning~\cite{swan2013quantified}. 
Besides predicting the action itself, it is also critical to predict its timing, so that recommendations and reminders can be made at the right time.
For instance, diet reminders ideally are delivered just \textit{before} meal choices are made~\cite{nahum2016just,thomas2015behavioral,freyne2017push}.
More generally, predicting user actions also enables digital personal assistants that support users with relevant information including local recommendations, traffic, weather, events, and news~\cite{du2016recurrent}. 

However, human behavior is extremely complex, which makes accurate predictions very challenging.
In particular, human behavior is (1) \textit{time-varying}, (2) \textit{interdependent}, and (3) \textit{periodic.}
First, real-world actions \textit{vary over time}, for example based on time of day (\eg, spending time with friends in the evenings) and day of week (\eg, going hiking on weekends)~\cite{koren2010collaborative,cheng2017troll}. 
Second, actions are also \textit{interdependent} in the short-term and the long-term (\eg, brushing teeth before going to bed, or drinking water after workouts). 
Third, humans are creatures of habit~\cite{davison1998predicting} 
and exhibit \textit{periodic} behaviors~\cite{das2012understanding,althoff2017harnessing,drutsa2017periodicity}, such as brushing teeth every morning and evening. 

Current user modeling techniques (\eg,~\cite{benson2016modeling,kapoor2015just,anderson2014dynamics,trouleau2016just,koren2010collaborative,davison1998predicting,gorniak2000predicting,zukerman1999predicting,lane1999hidden}) 
do not jointly model all these key aspects (time variation, interdependence, periodicity) of real-world action sequences.
However, failing to account for any of them results in decreased predictive performance.
For example, consider the task of predicting the time of a user's next meal. 
When not accounting for periodicity, one would miss the fact that the user's early lunch might lead to an earlier dinner as well. 
However, this could be a critical mistake if the user relies on timely diet reminders.

While great advances have been made in modeling specific aspects of behavior in narrow application domains, in particular in the space of recommender systems~\cite{koren2010collaborative} 
or information retrieval~\cite{agichtein2006improving,adar2008large,teevan2006history},
these lines of work have largely focused on consumption of items such as specific videos, songs, or websites~\cite{benson2016modeling,kapoor2015just,anderson2014dynamics,trouleau2016just,koren2010collaborative}. 
In all these cases, users repeat the \emph{same} high-level actions such as watching one video after another.
In contrast, we consider predicting \emph{which} higher-level action, out of many, the user will take next; 
for example, whether they will watch a movie or go for a run (not which specific movie or run). 
Furthermore, previous work has often focused on predicting short-term actions such as the next unix command~\cite{davison1998predicting}, web page request~\cite{zukerman1999predicting}, or TV episode watched~\cite{trouleau2016just}.
Instead, we are interested in predicting longer-term actions such as a commute in the evening or a run the next morning.

\xhdr{This work}
We present a new model for the task of predicting future user actions and their timing. 
First, we empirically demonstrate that action sequences exhibit time-varying, interdependent, and periodic patterns and that modeling them is critical to accurate predictions of user actions. 
Our model extends prior work on multivariate temporal point processes and is the first model to account for all three key properties. 
The model addresses 
(1)~time-varying propensities of actions through mixture of Gaussians,
(2)~short-term dependencies between actions through a Hawkes process, 
and (3)~long-term periodicity with time-dependent Weibull distributions. 
We call this model \emph{TIPAS} referring to Time-varying Interdependent Periodic Action Sequences. 
TIPAS is personalized to each user through learning user-specific action preferences.
We further develop an EM-based algorithm to fit this model using maximum likelihood estimation. 

We demonstrate that TIPAS can scale to real-world datasets from Argus and Under Armour activity logging applications that capture 12 million actions taken by 20 thousand users over 17 months. 
We evaluate our model on these two activity logging datasets capturing ten different real-world actions, and demonstrate that we can predict the user's next logged activity (\eg, run, eat, or sleep) and the timing of that activity (continuous, non-discretized timestamp)
based on the user's previous actions and their timing.

Further, we show that TIPAS accurately captures all three fundamental behavioral patterns in real-world data. 
Using several domains of real-world actions, we demonstrate that our model outperforms eleven
existing approaches on tasks of predicting actions by up to 156\%
as well as predicting when they will occur by up to 37\%.
Further, we show that performance improvements over baselines are particularly large for rare actions, increasing prediction accuracy over baselines by up to 256\%.
We find that these performance improvements are crucially enabled by modeling time-varying propensities of actions and their dependencies, and by modeling long-term periodicities of actions.
Empirically, modeling time-varying propensities of actions yields 53\% and 40\% accuracy on the two activity logging datasets.
Modeling short-term dependencies between actions improves this to 59\% and 49\%, respectively.
Also capturing long-term periodicities of actions further improves this to 61\% and 51\%, respectively.
Thus, capturing these three properties is essential to predicting periodic and interdependent human action sequences.

\section{Related Work}
\label{sec:related}

\xhdr{Predicting the next action}
Much work has focused on predictions of next actions, including 
unix commands ~\cite{davison1998predicting}, 
user interface actions to enable interface adaption~\cite{gorniak2000predicting},
web page requests 
allowing for prefetching and latency reduction~\cite{zukerman1999predicting}, 
clicks on web search~\cite{agichtein2006improving}, 
user behavior anomalies~\cite{lane1999hidden},
product item preferences \cite{koren2010collaborative,rendle2010factorizing}, 
online purchases~\cite{kooti2016portrait}, 
mobile apps used~\cite{baeza2015predicting},
and future location-based checkins~\cite{ashbrook2003using,liu2016predicting,bohnert2008using}.
Many of these works (\eg,~\cite{lane1999hidden,kapoor2015just,ashbrook2003using,bohnert2008using}) 
have formulated the problem as a discrete-time sequence prediction task and used Markov models.
However, Markov models assume unit time steps and are further unable to capture long-range dependencies since the overall state-space will grow exponentially in the number of time steps considered~\cite{du2016recurrent}.
Other works have used LSTM models~\cite{hochreiter1997long}, which also assume discrete time steps and are limited in their interpretability.

In contrast, we also model and predict \textit{when} the next action will occur,
which is critical to surface recommendations and reminders at the right time.
In addition, instead of specific web queries or item consumption, 
we consider a broader set of higher-level actions such as watching a movie, going for a run, or going to sleep.

\xhdr{Patterns of repeat consumption}
Another line of work has studied repeated actions, in particular in the space of item consumption, 
including video binge watching~\cite{trouleau2016just}, music listening~\cite{kapoor2015just}, web page revisitation patterns~\cite{adar2008large}, and repeated web search queries~\cite{teevan2006history}. 
More recent work has focused on modeling these behaviors and proposed models based on patterns of boredom~\cite{benson2016modeling,kapoor2015just} and recency~\cite{anderson2014dynamics}.

Importantly, patterns of human actions in the real world, which are modeled in this work, are fundamentally different from patterns of item consumption due to their higher-level notion (\eg, watching a movie, not which specific one). 
For example, patterns of boredom~\cite{benson2016modeling,kapoor2015just} suggest that the probability of repeating an action within a short amount of time is unlikely. 
In contrast, we empirically observe the opposite in some cases, such as users commuting one way being extremely likely to commute back in the near future.
More generally, real-world actions are characterized by more complex dynamics including time-varying behavior, interdependence, and periodicity of actions.

\xhdr{Temporal point processes}
Recent work has considered temporal point processes~\cite{cox1980point} including Poisson and Hawkes~\cite{hawkes1971spectra} process-based models to predict the timing of future actions.
Temporal point processes have been used to predict 
continuously time-varying item preferences~\cite{du2015time}, 
and to model user influence in a social network~\cite{iwata2013discovering,tanaka2016inferring,zhou2013learning}, 
the co-evolution of information and network structure~\cite{farajtabar2015coevolve}, 
competition between products~\cite{valera2015modeling}, 
mobility patterns in space and time~\cite{du2016recurrent}, 
user return times~\cite{kapoor2014hazard},
and temporal document clustering~\cite{du2015dirichlet,mavroforakis2016modeling}.
Perhaps the closest works to ours are by Du et al.~\cite{du2015time,du2016recurrent}, which also attempt to predict both future user actions \textit{and} their timing.
We extend this line of work by explicitly modeling time-varying action propensities as well as developing a novel combination of Exponential and Weibull kernels to model short-term and long-term periodic dependencies between actions. 
Further, we demonstrate that these aspects are critical when predicting real-world user actions and their timing across two real-world activity logging datasets.

\section{Task Description}
\label{sec:task}

%
%

The task considered in this work is, given a user and her history, a timestamped sequence of her actions in the past, 
to predict the user's future actions and the timing of these actions.

Formally, let $U$ be a set of users. 
Each user $u \in U$ has an action sequence, which we represent as a user history $H_{u} = \{(a_{un},t_{un})\}_{n=1}^{N_{u}}$ with a total of $N_{u}$ events.
Each element in $H_{u}$ is an event consisting of an action and timestamp representing that user $u$ takes action $a_{un} \in A$ at time $t_{un} \in \mathbb{R}^+$ ($ 0 \leq t_{un} \leq T$).
$T$ denotes the end of our observation period. 
For example, $a_{un}$ could correspond to watching a movie or going for a run (but not which specific movie or run).
We assume that events are sorted by their timestamps, $t_{un} \leq t_{un'}$ for $n < n'$.
We denote the set of events before time $t$ in user history $H_{u}$ as $H_{ut} = \{(a',t')|(a',t') \in H_{u}$ and $t' < t\}$. %

The task of predicting future user actions and when they will occur can now be formalized as follows.
Given user history $H_{ut}$ up until time $t$, predict the next $K$ actions the user will take and their timing $\{(a_k,t'_k)\}_{k=1}^K$, where $t'_k > t$ 
(\ie, these are the actions with the smallest $t'_k > t$ among all possible future user actions).

Here, we propose a novel multivariate temporal point process model for this prediction task and focus on the case of $K=1$.

\section{Empirical Observations}
\label{sec:observations}


Next we make a series of empirical observations about important properties of real-world action sequences that will provide the basis for our statistical model TIPAS (Section~\ref{sec:model}).
Accounting for these observations will lead to superior predictive models (Section~\ref{sec:experiments}).

\subsection{Dataset Description}\label{sec:argus_dataset_description}
To illustrate critical properties of real-world actions we use a dataset of logged activities from a mobile activity logging application, Argus by Azumio, used in previous work on activity logging~\cite{althoff2017onlineactions,althoff2017large,shameli2017gamification}. 
This smartphone app allows users to track their various daily activities including
drink, sleep, heart rate, running, weight, food, walking, biking, workout, and stretching actions.
For example, the drink action is logged to keep track of the user's daily fluid intake and the workout action is used to log various indoor exercises such as weightlifting or indoor-cycling.
This dataset includes over four thousand active users taking 1.2 million actions over the course of seven months (all users logged at least two unique actions per day on average). 
Due to the popularity of the app, this set of users is very diverse in terms of age, gender, health status, country of origin, and other features~\cite{althoff2017large}.
We note that the following properties of real-world actions also hold in other datasets including Under Armour activity logging app data (Section~\ref{sec:dataset}).

\subsection{Properties of Real-World Action Sequences}\label{sec:empirical_properties_overview}

Next, we describe three important properties of real-world action sequences and present empirical justification for each. 
TIPAS will explicitly address all three properties (Section~\ref{sec:model}). 
%

%
%

\begin{figure}[tbp]
  \centering
  \includegraphics[width=.80\columnwidth]{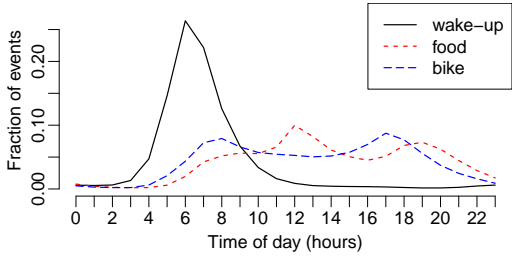}
  \caption{Fraction of events within each time-of-day window. 
  Notice that action propensity is clearly non-uniform and sometimes multi-modal. 
  }
  \vspace{-2mm}
  \label{fig:timeofday}
\end{figure}

\xhdr{Time-varying propensities of actions}
Human real-world actions vary over time, for example based on time of day (\eg, having meals in the morning, at mid-day, and in the evening) and day of week (\eg, working out on the weekends). 
This dynamic is evident in real-world data of human activities as illustrated in Figure~\ref{fig:timeofday}. 
The figure shows the distribution of the timing of three types of actions throughout the day: wake-up (from sleep), food, and bike. 
First, we observe that all three distributions are clearly non-uniform over time. 
For example, wake-up actions are clustered at around 07:00 hours (7 am). 
Second, we observe significant differences in the propensities to take different actions. 
While for sleep we observe a uni-modal distribution concentrated in the early morning, 
we observe a bi-modal distribution for biking. 
The two modes in the morning and evening likely correspond to commute activity where users log their rides to and from work. 
We also observe two clear modes for food during lunch and dinner times.
However, breakfast times seem to vary more widely across users and are more dispersed. 
Summarizing, we observe non-uniform, temporal distributions with varying number of modes that vary across actions.

\begin{figure}[tbp]
  \centering
  \includegraphics[width=41mm]{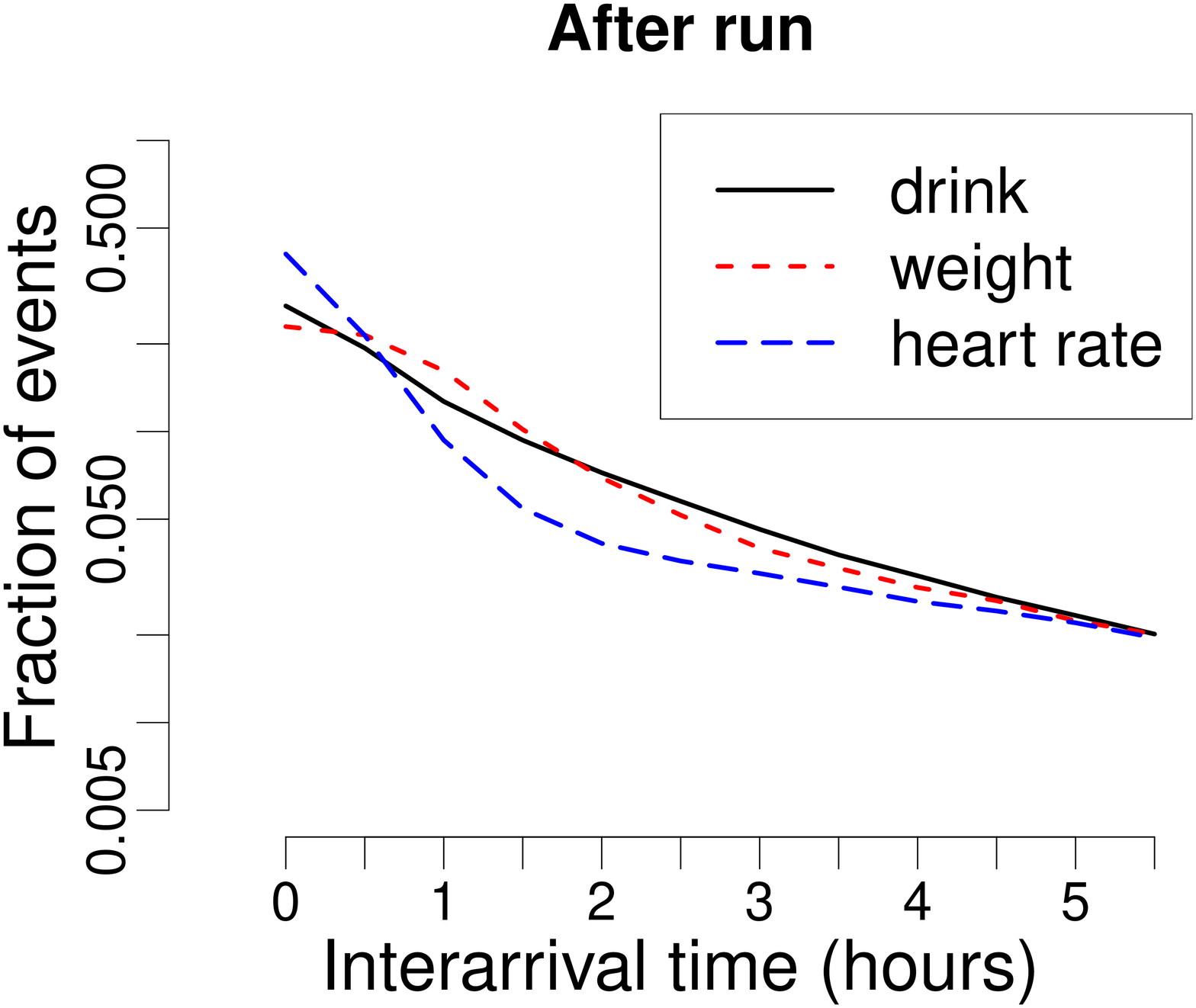}
  \hspace{1mm}
  \includegraphics[width=41mm]{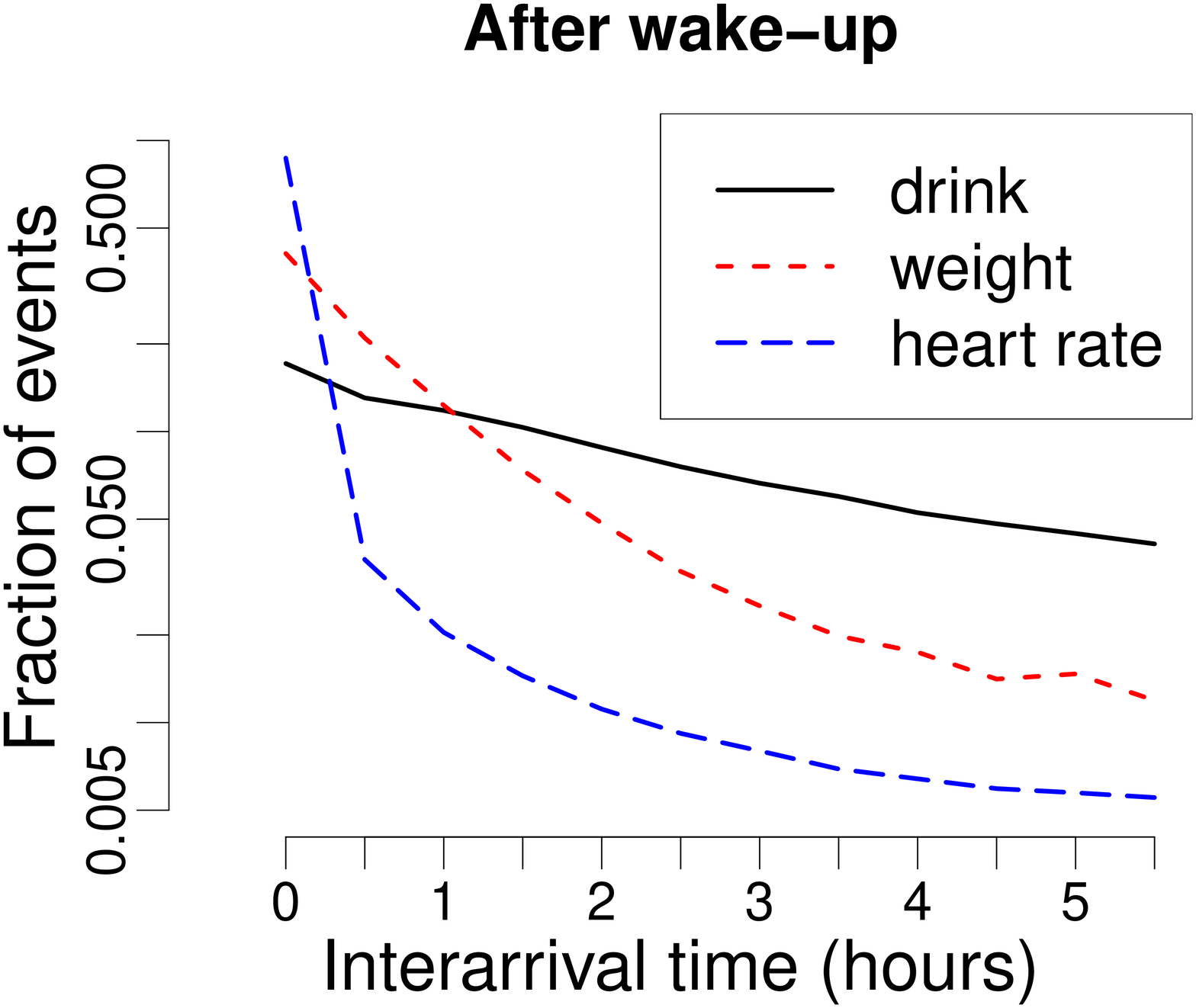}
  \caption{Fraction of interarrival times at each time window (log scale). 
  Figure shows drink, weight, and heart rate measurement actions taken after run (left) and wake-up (right) actions.
  Notice that the likelihood of drink, weight, and heart-rate actions declines quickly after both run and wake-up actions. 
  However, note that fraction of heart-rate actions decreases much quicker after wake-up than after runs. 
  \vspace{-2mm}
  }
  \label{fig:fraction_exponential}
\end{figure}

\xhdr{Short-term dependencies between actions}
Certain actions make it more likely that some other actions will follow shortly.
For example, people might drink water right after exercising or stretch right before running. 
In order to examine the short-term correlations between actions,
we extract interarrival times between pairs of actions (\ie, the elapsed time between the two actions) from a set of  action histories.
Figure~\ref{fig:fraction_exponential} shows the distribution of interarrival times for several pairs of user actions after run actions (left) and sleep actions (right).
We make two important observations.
First, the monotonically decreasing curves show that the likelihood of other actions is largest right after an action has happened. 
After this, the likelihood declines very quickly in a monotonic manner (note the log scale of the Y-axis). 
This points to a self-excitation dynamic of logged human actions.
For example, users are very likely to follow up on runs or waking up from sleep with drinking water or measuring their heart rate or weight. 
Specifically, about 50\% of the weight measurements which happen within 6 hours of waking up occur right within the first 30 minutes.  
Second, we find that the action dependency patterns vary across actions. 
For example, drinking is more common after runs than after waking up and heart rate measurements fall off more sharply right after waking up than after runs.
In summary, human actions in the real world often trigger other actions within a short period but these patterns are different across actions.
We can leverage these correlations among actions when predicting future events.


\begin{figure}[tbp]
  \centering
  \includegraphics[width=.88\columnwidth]{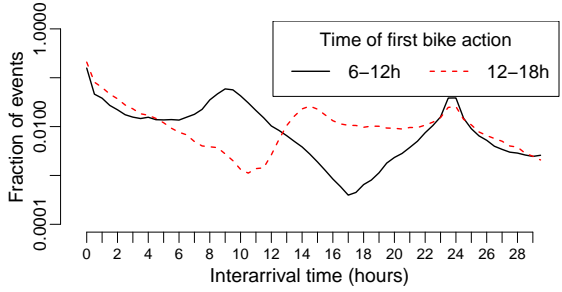} 
  \caption{
  Density describing when the next biking action will occur (interarrival time) given that the prior bike action occurred between 6-12h (solid black line) or between 12-18h (red dashed line) after midnight (timing, not duration). 
  Notice the multiple and different modes of the two distributions indicating that biking actions recur periodically but that the period timing depends on the time of day. 
  }
  \vspace{-2mm}
  \label{fig:fraction_rayleigh}
\end{figure}

\xhdr{Long-term periodic effects}
Humans exhibit periodic behaviors such as waking up at about the same time every morning or commuting back home after about 8 hours of work. 
Therefore, logged real-world actions likely follow periodic recurrence patterns in which the same action tends to recur at certain, regular intervals. 
While some of these periodic behaviors are rooted in intrinsic biological rhythms such as sleep~\cite{althoff2017harnessing}, others are dictated by extrinsic factors (\eg, when does one have to be in the office in the morning), or based on personal habits~\cite{davison1998predicting} (\eg, measuring one's weight before breakfast). 
We illustrate these dynamics using interarrival times between bike events in real-world data. 
Figure~\ref{fig:fraction_rayleigh} shows the distribution of interarrival times up to a maximum of 30h, where the two curves represent observed dynamics when the first of the two bike actions occurred during specific times of day (6-12h in solid black and 12-18h in dashed red line; note that these correspond to the timing and not the duration of the bike action).

We make two important observations.
Previously, we had observed that short-term dependencies between actions exhibit monotonic decay.
Here, we observe that this strong monotonic decay only holds within the first few hours and that we observe multiple additional peaks for both distributions after this initial phase.  
Second, we observe that these peaks occur at different times based on when the first action occurred.
In the case of the distribution for bike actions following a 6-12h bike ride, we observe peaks at around 9 and 24 hours (interarrival times), and peaks at around 14 and 24 hours for bike actions following a 12-18h bike ride. 
This behavior is not unexpected. When biking in the morning (6-12h), the next bike ride will likely be a commute back around 9h later. However, if the bike ride happens in the evening (12-18h), the next bike ride is likely not during the middle of the night, but after 14 hours or at around 8:00h in the morning. 
In addition, both curves exhibit a daily, 24h, periodicity. 
Modeling these periodicities allows us to capture user-specific timing of, for example, a late evening commute signaling a later start the next morning. 
In conclusion, two important dynamics could help predicting future real-world actions:
actions display periodic recurrence and the time of recurrence can depend on the time of day.

\vspace{-.5\baselineskip}

\section{Proposed Model}
\label{sec:model}

In this section, we operationalize the insights gained from empirical observations (Section~\ref{sec:observations}) in a probabilistic model based on temporal point processes, called TIPAS. 

\vspace{-.5\baselineskip}

\subsection{Background on Temporal Point Processes}\label{sec:math_background_temp_pp}

A temporal point process is a random process whose realization consists of a list of discrete events localized in time, $\{t_n\}_{n \in \mathbb{N}}$ with $t_n \in \mathbb{R}^+$. 
We introduce univariate temporal point processes for ease of exposition, though we will be using multivariate point processes to model the joint occurrence dynamics of multiple different actions 
(description inspired by~\cite{farajtabar2015coevolve}; more background in~\cite{aalen2008survival}). 
Let $H_t$ be the history of events before time $t$. 
Temporal point processes can be characterized via the conditional intensity function representing a stochastic model for the time of the next event given all the times of previous events. 
Formally, the conditional intensity function $\lambda(t)$ is the conditional probability of observing an event in a small window $[t, t + dt)$ given the history $H_t$; that is, $\lambda(t)dt = \mathbb{P}\{ \text{event in} [t, t + dt)|H_t\}.$ 
The conditional probability that no event happens during $[t, t')$ is $S(t') = \exp(-\int_t^{t'}\lambda(\tau)d\tau)$
and the conditional density that an event occurs at time $t'$ is $f(t') = \lambda(t') S(t')$~\cite{aalen2008survival}. 
Thus, the log-likelihood of a list of events $t_1, t_2, \dots , t_n$ in an observation window $[0, T)$, where $T>t_n$, can be expressed as 
\begin{align}
\mathcal{L}(t_1, t_2, \dots , t_n) = \sum_{i=1}^n {\log \lambda(t_i)} - \int_0^{T}\lambda(\tau)d\tau \, .
\label{eq:general_log_likelihood}
\end{align}
The intensity $\lambda$ can take various functional forms leading to a homogeneous Poisson process if $\lambda(t)$ is constant, to an inhomogeneous Poisson process if $\lambda(t)$ is time-varying but independent of the event history $H_t$, or to a Hawkes process if the intensity models mutual self-excitations between events~\cite{aalen2008survival}.
Our TIPAS model is based on multivariate Hawkes processes~\cite{hawkes1971spectra}.

\vspace{-.5\baselineskip}

\subsection{Model Definition}\label{sec:model_def}

\begin{figure}[tbp]
  \centering
  \includegraphics[width=0.90\columnwidth]{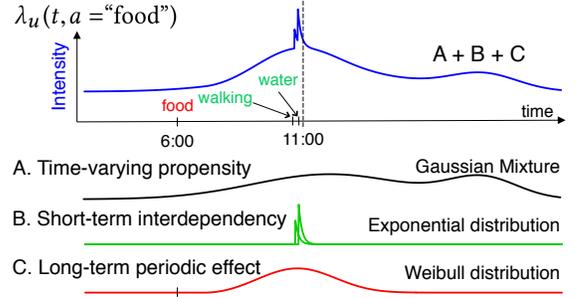} 
  \caption{
    Conceptual model overview.
    Intensity function of ``food'' for user $\bm{u}$ is modeled by the sum of three types of influences;
    time-varying background intensity (A; black), short-term dependencies (B; green), and long-term periodic effects (C; red).
    (A) Time-varying background intensity models typical times for food (\eg, having lunch around 12:00h). 
    (B) Events of ``walking'' and ``water'' might trigger ``food'' action within a short period of time.
    (C) Due to the early breakfast (6:00h), we might expect an earlier lunch.
  }
  \label{fig:model_visualization}
\end{figure}

We model user actions as a multivariate temporal point process with a time-varying intensity based on three factors based on our empirical observations (Section~\ref{sec:observations}).
The following intensity function models the rate that action $a$ occurs at time $t$ in user history $u$,
\begin{align}
  \lambda_{u}(t,a)\!=\!\alpha_{ua}\!\!+\!\!\mathit{Time}_{u}(t,a)\!\!+\!\!\mathit{ShortTerm}_{u}(t,a)\!\!+\!\!\mathit{LongTerm}_{u}(t,a). \!\!\label{eq:lambda}
\end{align}
Here, we use an additive decomposition of the intensity instead of modeling more complex interaction effects, 
because this approach is simple yet powerful and has been proven empirically successful as well~\cite{farajtabar2015coevolve,tanaka2016inferring,iwata2013discovering}. 
This model is conceptually visualized in Figure~\ref{fig:model_visualization}.
The figure shows how the overall intensity function $\lambda_{u}(t,a)$ (blue; here, $a=\text{food}$) is the sum of the time-varying propensity $\mathit{Time}_{u}(t,a)$ (black), short-term dependencies between actions \par \noindent
$\mathit{ShortTerm}_{u}(t,a)$ (green), and long-term periodic effects $\mathit{LongTerm}_{u}(t,a)$ (red) between actions
(for simplicity, we assume no personalization, \ie~$\alpha_{ua}=0$).
Note that our model does account for randomness, in the sense that not all actions may strictly conform to short-term and long-term patterns, through the personalized and time-varying baserates. 
In fact, learning model parameters from real data tries to account for all actions and will adapt distributional parameters to best explain all occurring actions.
Next, we formally define each of the four factors in turn.

\xhdr{Personalized action preferences: $\alpha_{ua}$}
We include personalized user preferences for specific actions through a constant additive factor $\alpha_{ua} \geq 0$ for each action and user.
Note that one could also model user preferences to be time-varying instead.
However, this would lead to a very large number of parameters and we show in Section~\ref{sec:experiments} that this simple model works well in practice. 

\xhdr{Time-varying propensities of actions: $\mathit{Time}_{u}(t,a)$}
Events can occur without influence from preceding events according to the background intensity function $\mathit{Time}_{u}(t,a)$. 
Having observed that the propensity of actions varies across time of day (Section~\ref{sec:empirical_properties_overview}),
we model the background intensity of action $a$ as a function of time-of-day through a Gaussian mixture model.
We define: 
\begin{eqnarray}
  \mathit{Time}_{u}(t,a) = \sum_{z \in \bm{Z}} \frac{\beta_{az}}{\sqrt{2\pi\sigma_{az}^{2}}}\exp\Bigl( - \frac{\bigl(l_{t}-\mu_{az}\bigr)^{2}}{2\sigma_{az}^{2}}\Bigr)\;\;,
  \label{eq:baserate}
\end{eqnarray}
where $z \in \bm{Z}$ represents the latent class of Gaussian mixture model (the number of mixtures can be determined through cross-validation).
For each action $a$ and latent mixture class $z$, $\mu_{az} > 0$ and $\sigma_{az} > 0$ denote the mean and standard deviation of the Gaussian distribution.
The importance of that mixture on the overall intensity function $\mathit{Time}_{u}(t,a)$ is captured by $\beta_{az} \geq 0$.
$l_{t}$ corresponds to the time of day of timestamp $t$ (\ie, elapsed time since midnight). 
We show in Section~\ref{sec:validating_parametric_assumptions} that Gaussian mixtures fit temporal variation in real-world data well.

\xhdr{Short-term dependencies between actions: $\mathit{ShortTerm}_{u}(t,a)$}
To model short-term dependencies between actions, we consider how the rate at which action $a$ occurs at time $t$ (Equation \ref{eq:lambda}) is influenced by actions $a'$ which occurred at previous time $t' < t$. 
We model these influences as a Hawkes process exhibiting self-excitations using Exponential decay functions starting at the time of previous actions. 
As demonstrated in Section~\ref{sec:empirical_properties_overview}, the short-term influence of previous actions diminishes quickly and monotonically, making the Exponential distribution a natural choice for the decay function. 
We define: 
\begin{align}
 \mathit{ShortTerm}_{u}(t,a) =  \sum_{(t',a') \in H_{ut} } \theta_{a'a}\omega_{a'a}\exp(-\omega_{a'a}\Delta_{t't}) \;\;,
   \label{eq:shortterm}
\end{align}
where $H_{ut} = \{(t',a')|(t',a') \in H_{u}$ and $t' < t\}$ is the set of events before time $t$ in history $u$,
and $\Delta_{t't} = t - t'$ is the time difference between time $t'$ and time $t > t'$. 
Further, $\omega_{a'a} \geq 0$ determines how quickly action $a'$ triggers action $a$ (shape of Exponential distribution),
and $\theta_{a'a} \geq 0$ determines how likely action $a'$ triggers action $a$ (scaling of distribution).
We estimate these parameters for each pair of actions $(a', a)$. 
Therefore, this component of the model captures the interdependencies between different actions (\eg, drinking after running), as well as the self-exciting effects of actions (\eg, running after running).
We show in Section~\ref{sec:validating_parametric_assumptions} that a Hawkes process with Exponential decay function fits short-term action dependencies in real-world data well.

\xhdr{Long-term periodic effects: $\mathit{LongTerm}_{u}(t,a)$}
We model the long-term periodic effects between identical actions (\eg, run to run) using Weibull distributions. 
The Weibull distribution is a continuous distribution with positive support (\ie, for $\Delta_{t't}>0$) that is well suited to model long-term effect patterns at different points in time and with different variance around its mean. 
We model the rate at which action $a$ occurs at time $t$
influenced by a previous event of action $a$ at time $t'$ as follows:
\begin{align}
  \!\!\!\! \mathit{LongTerm}_{u}(t,a) \! = \!\!\!\!\!\!\!\!\!\!\! \sum_{(t',a') \in H_{ut}^a}\!\!\!\!\!\!\!\!\! \phi_{c_{t'}a} \gamma_{c_{t'}a} \kappa_{c_{t'}a} \Delta_{t't}^{\kappa_{c_{t'}a}-1} \! \exp(- \gamma_{c_{t'}a} \Delta_{t't}^{\kappa_{c_{t'}a}}) \!\! \label{eq:longterm} 
\end{align}
where $H_{ut}^a = \{(t',a')|(t',a') \in H_{u}$ and $t' < t$ and $a' = a\}$ 
is the set of events of action $a$ before time $t$ in history $u$,
and $\Delta_{t't} = t - t'$ is again the time difference between time $t'$ and time $t>t'$.
As shown in Section~\ref{sec:empirical_properties_overview}, long-term effects vary based on the time of day of action $a'$. 
This is captured through the parameter $c_{t'}\in C$ that represents discretized time-of-day windows (\eg, using four classes as 0-6h, 6-12h, 12-18h, and 18-24h). 
This allows us to learn time-of-day-dependent distributions modeling different periodicities. 
Parametrized by this time-of-day category $c_{t'}$ and by action $a$,
$\gamma_{c_{t'}a} \geq 0$, $\phi_{c_{t'}a} \geq 0$ determine how quickly and how likely (influence)
action $a'$ (which occurred in time-of-day window $c_{t'}$) triggered its following event of action $a$.
$\kappa_{c_{t'}a} \geq 0$ determines the shape of the Weibull distribution.
In Section~\ref{sec:validating_parametric_assumptions}, we demonstrate that the Weibull distribution closely match periodic dynamics in real-world data.

\subsection{Model Inference}\label{sec:inference}
We use maximum likelihood estimation to infer the parameters of our proposed model (Equation~\ref{eq:lambda}).
The unknown parameters are 
$\bm{\alpha} = \{\{\alpha_{ua}\}_{u \in U}\}_{a \in A}$,
$\bm{\beta} = \{\{\beta_{az}\}_{a \in A}\}_{z \in \bm{Z}}$,
$\bm{\mu} = \{\{\mu_{az}\}_{a \in A}\}_{z \in \bm{Z}}$,
$\bm{\sigma} = \{\{\sigma_{az}\}_{a \in A}\}_{z \in \bm{Z}}$,
$\bm{\Theta} = \{\{\theta_{a'a}\}_{a \in A}\}_{a' \in A}$,
$\bm{\Omega} = \{\{\omega_{a'a}\}_{a \in A}\}_{a' \in A}$,
$\bm{\Phi} = \{\{\phi_{ca}\}_{c \in C}\}_{a \in A}$,
$\bm{\Gamma} = \{\{\gamma_{ca}\}_{c \in C}\}_{a \in A}$,
and $\bm{K} = \{\{\kappa_{ca}\}_{c \in C}\}_{a \in A}$.
The set of all parameters is denoted by $\bm{\Psi}=\{\bm{\alpha}, \bm{\beta}, \bm{\mu}, \bm{\sigma}, \bm{\Theta}, \bm{\Omega}, \bm{\Phi}, \bm{\Gamma}, \bm{K}\}$.

The log-likelihood function (Equation~\ref{eq:general_log_likelihood}), given a set of user histories $\mathcal{H} = \{H_u\}_{u \in U}$, can be expressed as: 
{\small
\vspace{-.25\baselineskip}
\begin{align}
  \mathcal{L}(\bm{\Psi}|\mathcal{H}) = \sum_{u \in U}\sum_{n=1}^{N_{u}} \log \lambda_{u}(t_{un},a_{un}) - \sum_{u \in U} \int_{0}^{T} \sum_{a \in A} \lambda_{u}(t,a)dt \;\;,
  \label{eq:loglikelihood1}
\end{align}
}
where the last term, the expectation function, represents the expected number of events in the time period from 0 to $T$. 
Combining Equations (\ref{eq:lambda})-(\ref{eq:loglikelihood1}),
the log-likelihood can be written as follows:
{\small
\vspace{-.25\baselineskip}
\begin{align}
  \lefteqn{\mathcal{L}(\bm{\Psi}|\mathcal{H}) = }  \nonumber \\[-3.0pt]
  & \sum_{u \in U} \sum_{n=1}^{N_{u}} \log \Biggl\{ \alpha_{ua_{un}} + \sum_{z \in \bm{Z}} \frac{\beta_{a_{un}z}}{\sqrt{2\pi\sigma_{a_{un}z}^{2}}}\exp\Bigl( - \frac{\bigl(l_{t_{un}}-\mu_{a_{un}z}\bigr)^{2}}{2\sigma_{a_{un}z}^{2}}\Bigr) \nonumber \\[-3.0pt]
  & + \sum_{m=1}^{n-1} \theta_{a_{um}a_{un}} \omega_{a_{um}a_{un}} \exp (-\omega_{a_{um}a_{un}} \Delta_{t_{um}t_{un}})\nonumber \\[-3.0pt]
  & + \sum_{l=1}^{n-1} \Bigl( I(a_{ul}=a_{un}) \phi_{c_{ul}a_{un}} \gamma_{c_{ul}a_{un}} \kappa_{c_{ul}a_{un}} \Delta_{t_{ul}t_{un}}^{\kappa_{c_{ul}a_{un}}-1}  \nonumber \\[-3.0pt]
  & \times \exp( -\gamma_{c_{ul}a_{un}} \Delta_{t_{ul}t_{un}}^{\kappa_{c_{ul}a_{un}}} ) \Bigr) \Biggr\} - \sum_{u \in U} \int_{0}^{T} \sum_{a \in A} \lambda_{u}(t,a)dt \;\;,
  \label{eq:loglikelihood2}
\end{align}
}
where $c_{ul} \in C$ represents time-of-day category of $l$-th event of $u$,
and $I(\cdot)$ is the indicator function.
The integral in Equation~\ref{eq:loglikelihood2} can be analytically calculated.

Inspired by previous work~\cite{zhou2013learning,farajtabar2015coevolve}, we develop an efficient inference algorithm to maximize the log-likelihood based on the EM algorithm. 
By iterating the E-step and the M-step until convergence,
we obtain a local optimum solution for $\bm{\Psi}$.

\xhdr{E-step}
Conceptually, we introduce latent variables $\bm{p},\bm{q},\bm{r}$ to capture why each event was triggered either through user preference, time-varying background intensity, short-term action interdependencies, or long-term periodic effects.
Let $p_{0,un}$ be the probability that the $n$-th event of user $u$ was triggered by user preference,
$p_{z,un}$ be the probability that the $n$-th event of user $u$ was triggered by the time-varying background intensity function of latent class $z$,
$q_{um,un}$ be the probability that the $n$-th event of user $u$ was triggered by the short-term effect of the $m$-th event of user $u$,
and $r_{ul,un}$ be the probability that the $n$-th event of user $u$ was triggered by the long-term effect of the $l$-th event of user $u$.

In E-step, $k$-th estimate of $p^{k}_{0,un}$, $p_{z,un}^{k}$, $q^{k}_{um,un}$, and $r^{k}_{ul,un}$ are calculated by: 
{\small
\vspace{-.25\baselineskip}
\begin{align}
  p^{k}_{0,un} = \frac{\alpha^{k}_{ua_{un}}}{R_{un}} \;\;,
\end{align}
\vspace{-.75\baselineskip}
\begin{align}
  p_{z,un}^{k} = \frac{1}{R_{un}} \frac{\beta_{a_{un}z}^{k}}{\sqrt{2\pi(\sigma_{a_{un}z}^{k})^{2}}}\exp\Bigl( - \frac{\bigl(l_{t_{un}}-\mu^{k}_{a_{un}z}\bigr)^{2}}{2(\sigma^{k}_{a_{un}z})^{2}}\Bigr) \;\;,
\end{align}
\vspace{-.75\baselineskip}
\begin{align}
  q^{k}_{um,un} = \frac{1}{R_{un}} \theta^{k}_{a_{um}a_{un}} \omega^{k}_{a_{um}a_{un}} \exp (-\omega^{k}_{a_{um}a_{un}} \Delta_{t_{um}t_{un}}) \;\;,
\end{align}
\vspace{-.75\baselineskip}
\begin{align}
  \lefteqn{ r^{k}_{ul,un} = \Biggl\{ \frac{1}{R_{un}} \phi^{k}_{c_{ul}a_{un}} \gamma^{k}_{c_{ul}a_{un}} \kappa^{k}_{c_{ul}a_{un}} } \nonumber \\[-3.0pt]
  & \quad \qquad \qquad \times \Delta_{t_{ul}t_{un}}^{\kappa^{k}_{c_{ul}a_{un}} - 1} \exp(- \gamma^{k}_{c_{ul}a_{un}} \Delta_{t_{ul}t_{un}}^{\kappa^{k}_{c_{ul}a_{un}}}) \Biggr\}\;\;,
\end{align}
}
where $\bm{\Psi}^{k}=\{\bm{\alpha}^{k}, \bm{\beta}^{k}, \bm{\mu}^{k}, \bm{\sigma}^{k}, \bm{\Theta}^{k}, \bm{\Omega}^{k}, \bm{\Phi}^{k}, \bm{\Gamma}^{k}, \bm{K}^{k}\}$ is the $k$-th estimate of parameters in the EM procedure,
and $R_{un}$ is the normalization factor in order to satisfy $p^{k}_{0,un} + \sum_{z \in \bm{Z}} p^{k}_{z,un} + \sum_{m=1}^{n-1} q^{k}_{um,un} + \sum_{l=1}^{n-1} r^{k}_{ul,un} = 1$.

\xhdr{M-step}
We use Jensen's inequality to provide a lower bound for the log-likelihood (Equation~\ref{eq:loglikelihood2}); this lower bound is often called the $Q$ function.
We obtain the next estimate of the parameters by taking the derivative of the $Q$ function with respect to each parameter and setting them to zero: 
{\small
\vspace{-.25\baselineskip}
\begin{align}
  \alpha_{ua}^{k+1} = \frac{\sum_{n=1}^{N_{u}}I(a_{un}=a)p_{0,un}^{k}}{T} \;\;,
\end{align}
\vspace{-.75\baselineskip}
\begin{align}
  \beta_{az}^{k+1} = \frac{2 \mathcal{T}}{|U| T} \times \frac{\sum_{u \in U} \sum_{n=1}^{N_{u}} I(a_{un}=a) p_{z,un}^{k}}
       { \mathit{erf}(\frac{\mu_{az}^{k}}{\sqrt{2}\sigma_{az}^{k}}) + \mathit{erf}(\frac{\mathcal{T} - \mu_{az}^{k}}{\sqrt{2}\sigma_{az}^{k}}) } \;\;,
\end{align}
\vspace{-.75\baselineskip}
\begin{align}
  \theta_{a'a}^{k+1} = 
\frac{\sum_{u \in U} \sum_{n=1}^{N_{u}} \sum_{m=1}^{n-1} I(a_{um}=a',a_{un}=a) q^{k}_{um,un}}
{\sum_{u \in U} \sum_{n=1}^{N_{u}} I(a_{un}=a') \Bigl( 1 - \exp \bigl( - \omega_{a'a}^{k} (T-t_{un}) \bigr) \Bigr)} \;\;,
\end{align}
\vspace{-.75\baselineskip}
\begin{align}
  \phi_{ca}^{k+1} = 
  \frac{\sum_{u \in U} \sum_{n=1}^{N_{u}} \sum_{l=1}^{n-1} I(a_{ul}=a,a_{un}=a,c_{ul}=c) r^{k}_{ul,un}}
       {\sum_{u \in U} \sum_{n=1}^{N_{u}} I(a_{un}=a,c_{un}=c) \Bigl( 1 - \exp \bigl( -\gamma_{ca}^{k}(T-t_{un})^{\kappa_{ca}^{k}} \bigr) \Bigr)} \;\;,
\end{align}
}
where $\mathcal{T}$ is the time period of a day (\ie, 24 hours),
$\frac{T}{\mathcal{T}}$ is the number of days representation of the observed period $T$,
and where $\mathit{erf}$ denotes the Gauss error function $\mathit{erf}(x) = \frac{1}{\sqrt\pi}\int_{-x}^x e^{-t^2} \,\mathrm dt $.
Because of the exponentials ($\exp$ and $\mathit{erf}$) within the expectation function (Equation~\ref{eq:loglikelihood2}), 
$\omega_{a'a}^{k+1}$, $\gamma_{ca}^{k+1}$, $\kappa_{ca}^{k+1}$, $\mu^{k+1}_{az}$, and $\sigma^{k+1}_{az}$ cannot be solved in closed form. 
However, by further considering a lower bound for these exponentials $\omega_{a'a}^{k+1}$ and $\gamma_{ca}^{k+1}$ can be solved in closed form.
Their update rules are as follows:
{\small
\vspace{-.25\baselineskip}
\begin{align}
  \lefteqn{ \omega_{a'a}^{k+1} = \Biggl\{
    \sum_{u \in U} \sum_{n=1}^{N_{u}} \sum_{m=1}^{n-1} I(a_{um}=a',a_{un}=a) q^{k}_{um,un}
    \Biggr\} } \nonumber \\[-3.0pt]
  & / \Biggl\{ \sum_{u \in U} \sum_{n=1}^{N_{u}} \sum_{m=1}^{n-1} I(a_{um}=a',a_{un}=a) q^{k}_{um,un} \Delta_{t_{um}t_{un}} \nonumber \\[-3.0pt]
  & + \sum_{u \in U} \sum_{n=1}^{N_{u}} I(a_{un}=a') \theta_{a'a}^{k} (T-t_{un})\exp\bigl(-\omega_{a'a}^{k}(T-t_{un})\bigr) \Biggr\} \;\;,
\end{align}
\vspace{-1\baselineskip}
\begin{align}
  \lefteqn{ \gamma_{ca}^{k+1} = \Biggl\{ \sum_{u \in U} \sum_{n=1}^{N_{u}} \sum_{l=1}^{n-1} I(a_{ul}=a,a_{un}=a,c_{ul}=c) r^{k}_{ul,un} \Biggr\} } \nonumber \\[-3.0pt]
  & / \Biggl\{ \sum_{u \in U} \sum_{n=1}^{N_{u}} \sum_{l=1}^{n-1} I(a_{ul}=a,a_{un}=a,c_{ul}=c) r^{k}_{ul,un} \Delta_{t_{ul}t_{un}}^{\kappa^{k}_{ca}} \nonumber \\[-3.0pt]
  & + \sum_{u \in U} \sum_{n=1}^{N_{u}} I(a_{un}=a, c_{un}=c) \phi^{k}_{ca} ( T - t_{un})^{\kappa^{k}_{ca}} \exp\bigl( - \gamma_{ca}^{k} (T-t_{un})^{\kappa^{k}_{ca}}\bigr) \Biggr\} \;.
\end{align}
}
The other three parameters, $\kappa_{ca}^{k+1}$, $\mu^{k+1}_{az}$ and $\sigma^{k+1}_{az}$, are estimated by maximizing the $Q$ function through the use of a gradient-based numerical optimization method; we used the Newton method.
For more details on model inference see the Online Appendix~\cite{kurashima2017onlineappendix}.

\section{Experiments}
\label{sec:experiments}

This section evaluates the predictive performance of our proposed model on two real-world datasets on predicting the next user action and when it will occur.
We compare against eleven different baselines on each dataset.
However, since many baseline models are unable to make joint predictions of action and timing,
we evaluate these two tasks separately.
Importantly, this process allows us to identify the individual sources of error that would impact joint predictions.
Our implementation is available at \url{snap.stanford.edu/tipas}.
\vspace{-\baselineskip}

\subsection{Datasets}\label{sec:dataset}

Our experiments use two real-world activity logging datasets.
In total, these datasets comprise 12 millions real-world actions taken by 20 thousand users over 17 months.

\begin{table}[tbp]
\centering
\resizebox{.99\columnwidth}{!}{%
\begin{tabular}{lrr}
  \toprule
  Dataset Statistics & Argus & Under Armour \\ \midrule
  Observation period & 7 months & 10 months \\
                     & Jan-July '15 & Jan-Oct '16 \\
  \# unique actions & 10 & 8\\
  \# total users & 4,708 & 15,221 \\ 
  \# total actions & 2,140,757 & 9,733,645 \\
  Avg. \# actions per user & 454.7 & 639.5\\
  Avg. \# unique actions per user & 6.3 & 6.8 \\
  Avg. \# unique actions per user day & 2.7 & 4.4 \\
  \bottomrule
 \end{tabular}
 }
 \caption{Basic dataset statistics.}
 \label{tab:dataset}
  \vspace{-3mm}
\end{table}

\xhdr{Argus dataset}
We use the activity logging data from the Argus mobile app described in Section~\ref{sec:argus_dataset_description}.
Users in this dataset can log 10 different actions (drink, sleep, heart rate, running, weight, food, walking, biking, workout, and stretching) and our goal is to predict which of these 10 actions a user will take next (and when).
Our analyses include users who logged at least two unique actions per day on average (other users might only use the app to for example track their sleep making predictions of actions and their timing almost trivial; we find that our results are robust to different choices of this threshold).  
We consider 7 months of data from the app in a rolling window evaluation, where we use one month for training and the next for testing (\ie, making out-of-sample predictions; without retraining).
As shown in Table~\ref{tab:dataset}, the dataset includes 2.1 million actions by over 4 thousand users within the 7 month observation period.

\xhdr{Under Armour dataset (UA)}
We also use activity logging data from Under Armour mobile apps (\ie, MapMyFitness and MyFitnessPal; focusing on users that are active in both apps). 
Users in this dataset can log 8 different types of actions (running, walking, biking, workout, breakfast, lunch, dinner, and snacks).
Our analyses include users who logged at least four unique actions per day on average, leading to a similar number of unique actions per user on average compared to the Argus dataset (again, our results are robust to different choices of this threshold).
We consider 10 months of data from the app and again perform a rolling window evaluation where we train on one month and test on the next.
In total, this dataset comprises 15 thousand users taking 9.8 million actions (Table~\ref{tab:dataset}).

\subsection{Model Learning}\label{sec:model_learning}

Note that our model has few core model parameters.
In the context of the datasets described above, we have about 500 core model parameters ($\bm{\beta}, \bm{\mu}, \bm{\sigma}, \bm{\Theta}, \bm{\Omega}, \bm{\Phi}, \bm{\Gamma}, \bm{K}$) and about 25 thousand personalization parameters ($\bm{\alpha}$).
This small, non-redundant set of parameters allows us to train the model efficiently and robustly, and explain model predictions through inspection and visualization of model parameters (Section~\ref{subsec:interpretability}), while performing competitively (Section~\ref{sec:behaviorprediction}).
However, during training time (but not test time) we also have latent variables ($\bm{p},\bm{q},\bm{r}$) that allow us to learn the core model parameters.
These latent variables represent which actions trigger which other actions, leading to $\mathcal{O}(|U| (max_{u \in U} N_u)^2)$ variables in the worst case.
On both datasets, inference of both core model and latent parameters involves solving an optimization problem with over 200 million total variables (Section~\ref{sec:inference}; we randomly initialize all parameters).
Using our EM-based inference procedure we can robustly infer these parameters in less than ten hours using a single-threaded C++ implementation on a single machine. 
We find that one month of training data is enough to reliably train our model.

\subsection{Validating Parametric Assumptions}\label{sec:validating_parametric_assumptions}

\begin{figure}[tbp]
  \centering
  \vspace{-4mm}
  \subfigure[Time-varying propensity (bike)]{\label{fig:validating_parametric_assumptions_a}\centering\includegraphics[width=.48\columnwidth]{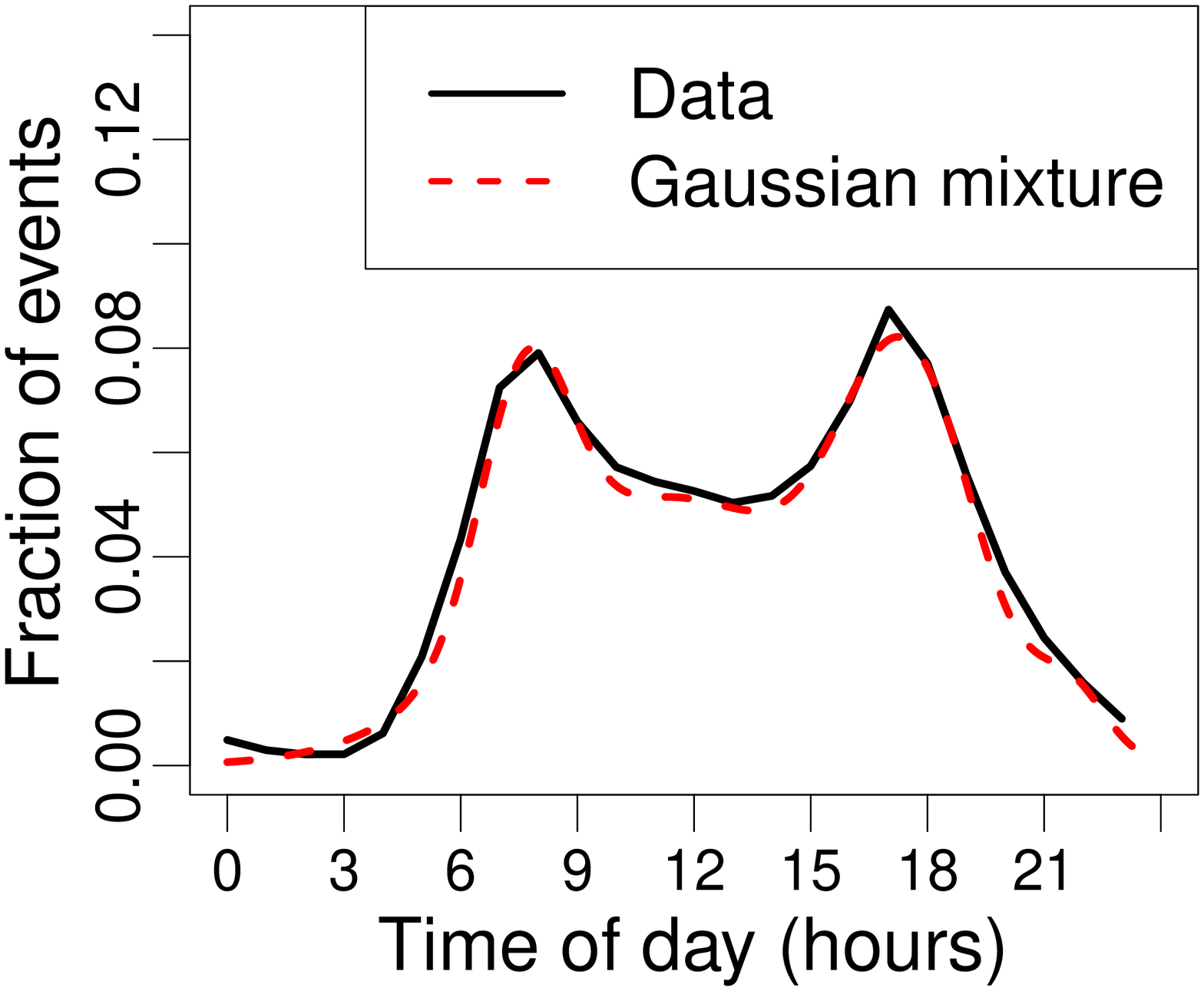}}
  \hspace{1mm}
  \subfigure[Short- \& long-term effects (bike)]{\label{fig:validating_parametric_assumptions_b}\centering\includegraphics[width=.48\columnwidth]{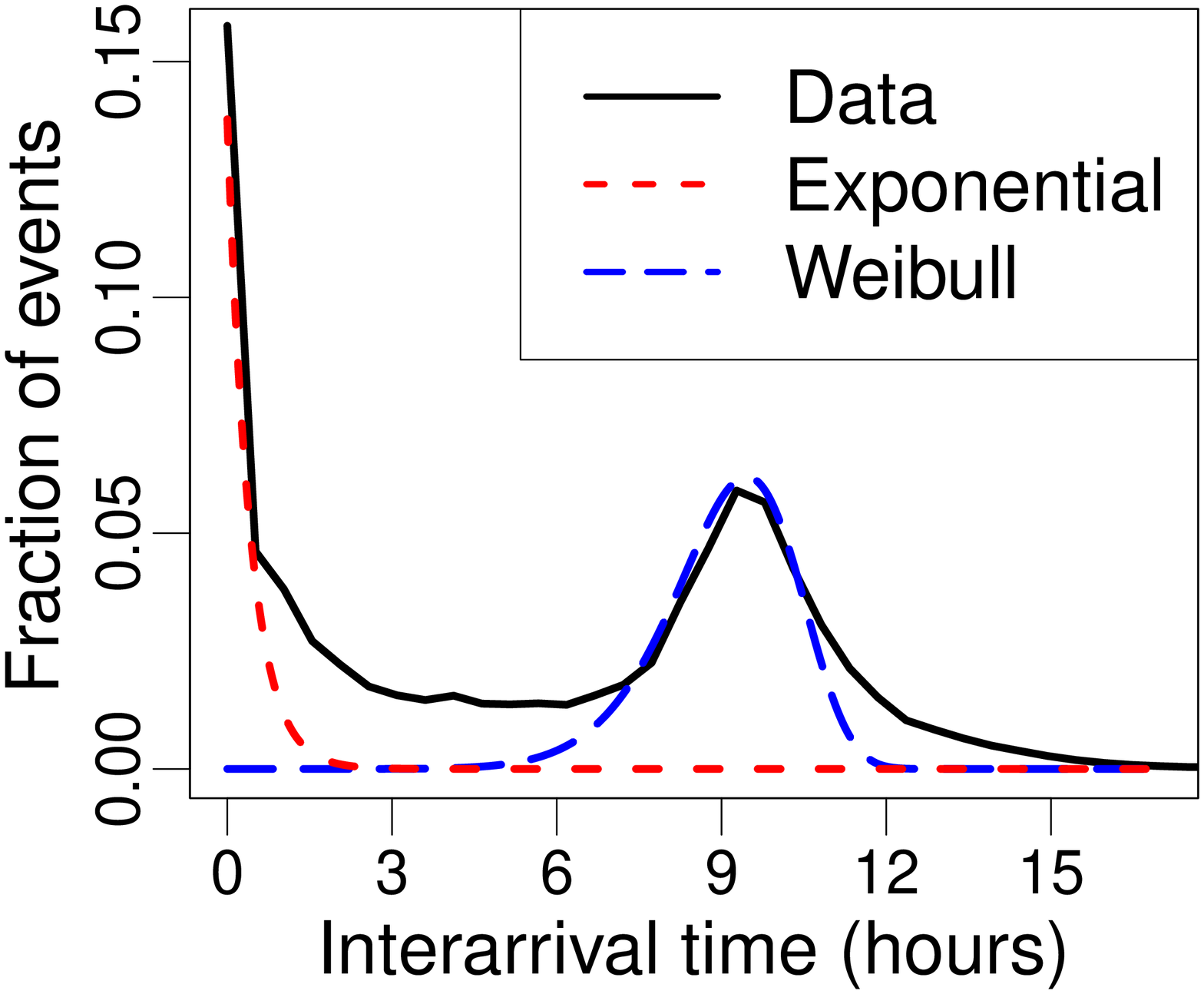}}
  \vspace{-2mm}
  \caption{
  Validation of parametric modeling assumptions (Section~\ref{sec:model_def}). (a) Mixture of Gaussian closely fits observed time-varying action propensity (here, for bike action). (b) Exponential and Weibull distributions collectively well-approximate short-term dependencies and long-term periodic effects of previous bike actions.
  }
  \vspace{-2mm}
  \label{fig:validating_parametric_assumptions}
\end{figure}

In Section~\ref{sec:model_def} we developed a model consisting of three parts: time variation modeled using a mixture of Gaussians ($\mathrm{Time}_{u}(t,a)$), short-term dependencies between actions modeled by a Hawkes process with Exponential decay function ($\mathrm{ShortTerm}_{u}(t,a)$), and long-term periodicity modeled through Weibull distributions ($\mathrm{LongTerm}_{u}(t,a)$).
Here, we test empirically whether these parametric assumptions hold true in real data. 
Using the Argus dataset, we inferred appropriate parameters for these distributions.

We demonstrate qualitatively in Figure~\ref{fig:validating_parametric_assumptions} that the chosen distributions fit real-world dynamics well. 
Figure~\ref{fig:validating_parametric_assumptions} (a) shows the time-varying propensity with superimposed mixture of Gaussian fit and
(b) shows that, collectively, Exponential and Weibull distribution closely approximate the influence of previous actions (example data is for bike action as seen in Figure~\ref{fig:timeofday}).

We have further quantitatively evaluated our parametric assumptions and compared our choices to alternative distributions (\eg, Rayleigh and Power-law) through goodness-of-fit tests which have shown that the suggested distributions best fit real-world data.

\subsection{Predicting the Next Action}\label{sec:behaviorprediction}

First, we evaluate our proposed model in terms of its accuracy in predicting actions at a given time. 
The task is to predict the \mbox{$n\!+\!1$-st} action $a_{u n+1}$ of user $u$,
given time $t_{u n+1}$ and past user history $\bm{H}_{u}=\{(a_{u1},t_{u1}),\cdots,(a_{un},t_{un})\}$.
For each two month period in both datasets, we use the first month for training and the second month for testing and perform a rolling window evaluation, where we predict each test set event given all events that happened before it (without retraining). 
We use accuracy, the percentage of correct predictions, over all test events as our evaluation measure (the most common measure to evaluate recommender systems~\cite{herlocker2004evaluating}).
We also report macro-averaged recall~\cite{manning2008introduction} corresponding to averaging prediction accuracy equally weighted across all action types. 
This measure highlights differences in predictive performance on rare actions that do not affect the standard accuracy measure very much. 
We find very similar results using other classification metrics (\eg, ROC AUC, F1). 
The number of mixtures for the time-varying action propensity (Equation~\ref{eq:baserate}) is set via cross-validation. 
We compare our proposed model against the following seven 
baseline models, which have proven competitive across a wide variety of prediction tasks and recommender systems:
\begin{itemize}
\item \textbf{Copy Model:} Simply repeats the user's last action.
  Several repeat consumption models are variants of this copy model (\eg,~\cite{anderson2014dynamics,benson2016modeling}).
\item \textbf{Markov Model:} Predicts the next action based on the most recent actions of the user.
We report first to fifth-order Markov models (sixth-order models did not significantly improve performance).
Markov models have been used widely to predict next actions (\eg,~\cite{ashbrook2003using,kapoor2015just}).
\item \textbf{Hidden Markov Model (HMM):} This is a Markov model with hidden (unobserved) states. It predicts the next action based on the current, inferred state of the action sequence \cite{lane1999hidden}.
\item \textbf{Factorizing Personalized Markov Chains (FPMC):} This is based on underlying Markov chains where the transitions matrices are user-specific. Matrix factorization models are used to address sparsity issues of these user-specific Markov chains~\cite{rendle2010factorizing}.
\item
\textbf{Recurrent Neural Network (RNN):} Feedforward neural network structure using outputs from the hidden units at the prior time step as the inputs as the current time step.
Assumes discrete time steps and no ready-to-use generalizations to continuous time domain exist.
\item \textbf{PP-Global:} A global Poisson process model. The intensity function is constant over time and defined by $\lambda_{u}(t,a) = \alpha_{a}$. 
\item \textbf{PP-User:} A user-specific Poisson process model. The intensity function is constant over time and defined by $\lambda_{u}(t,a) = \alpha_{ua}$. 
\end{itemize}
Note that Hawkes process models (\eg,~\cite{hawkes1971spectra,farajtabar2015coevolve,du2015dirichlet}) are closely related to the $\mathrm{ShortTerm}_{u}(t,a)$ component of our model (Equation~\ref{eq:shortterm}). 
Our proposed model \textbf{TIPAS}
uses the intensity function of Eq.~\ref{eq:lambda}.
We predict the most likely user action as $\hat a = \argmax_a \lambda_{u}(t_{u n+1},a)$. 
We also compare the individual model components in an ablation study below.

\begin{figure}[tbp]
  \centering
  \vspace{-1mm}
  \subfigure[Baseline Comparison (Argus)]{\label{fig:action_prediction_micro_baselines_a}\centering\includegraphics[width=.44\columnwidth]{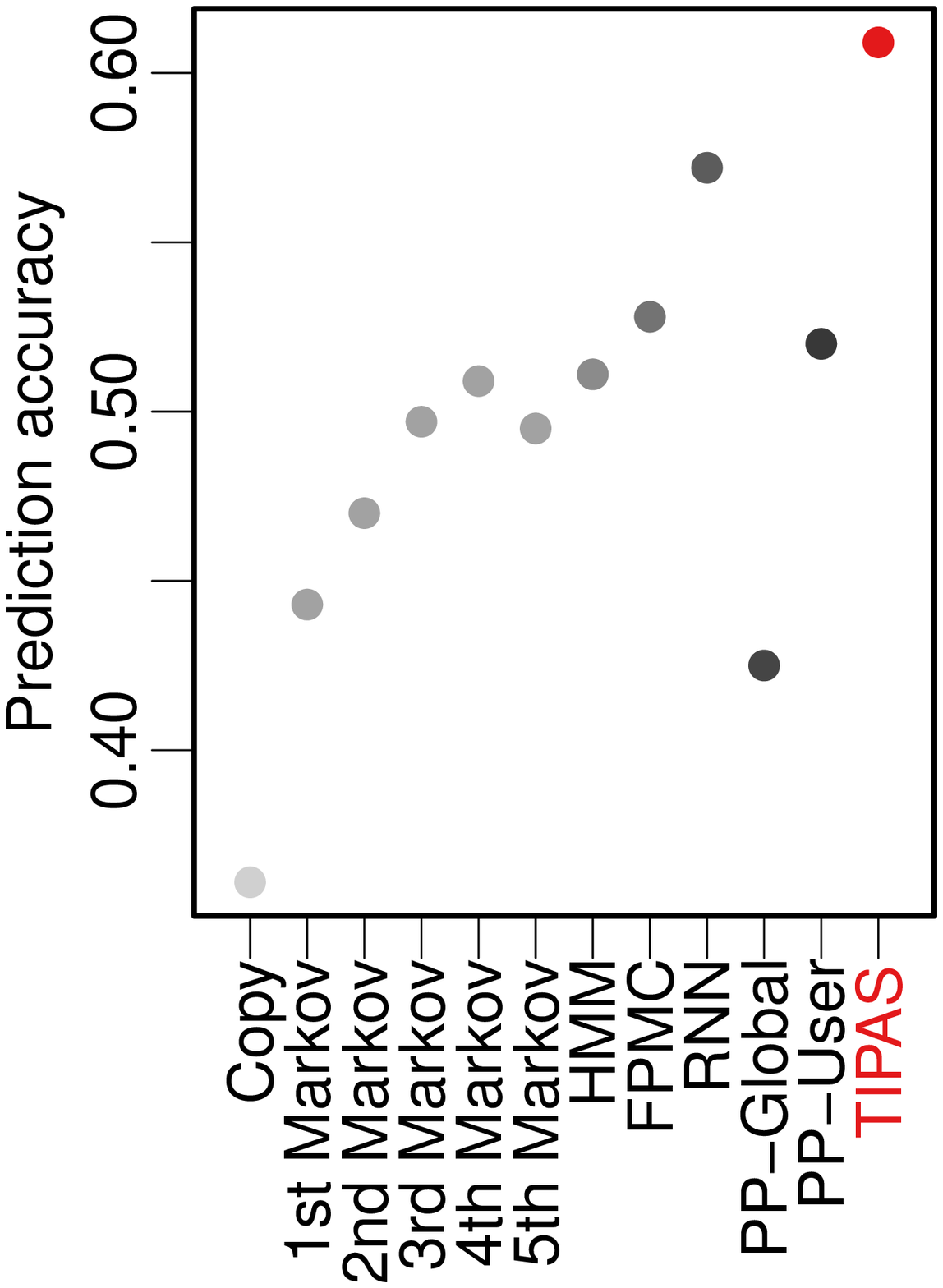}} 
    \hspace{5mm}
  \subfigure[Baseline Comparison (UA)]{\label{fig:action_prediction_micro_baselines_b}\centering\includegraphics[width=.44\columnwidth]{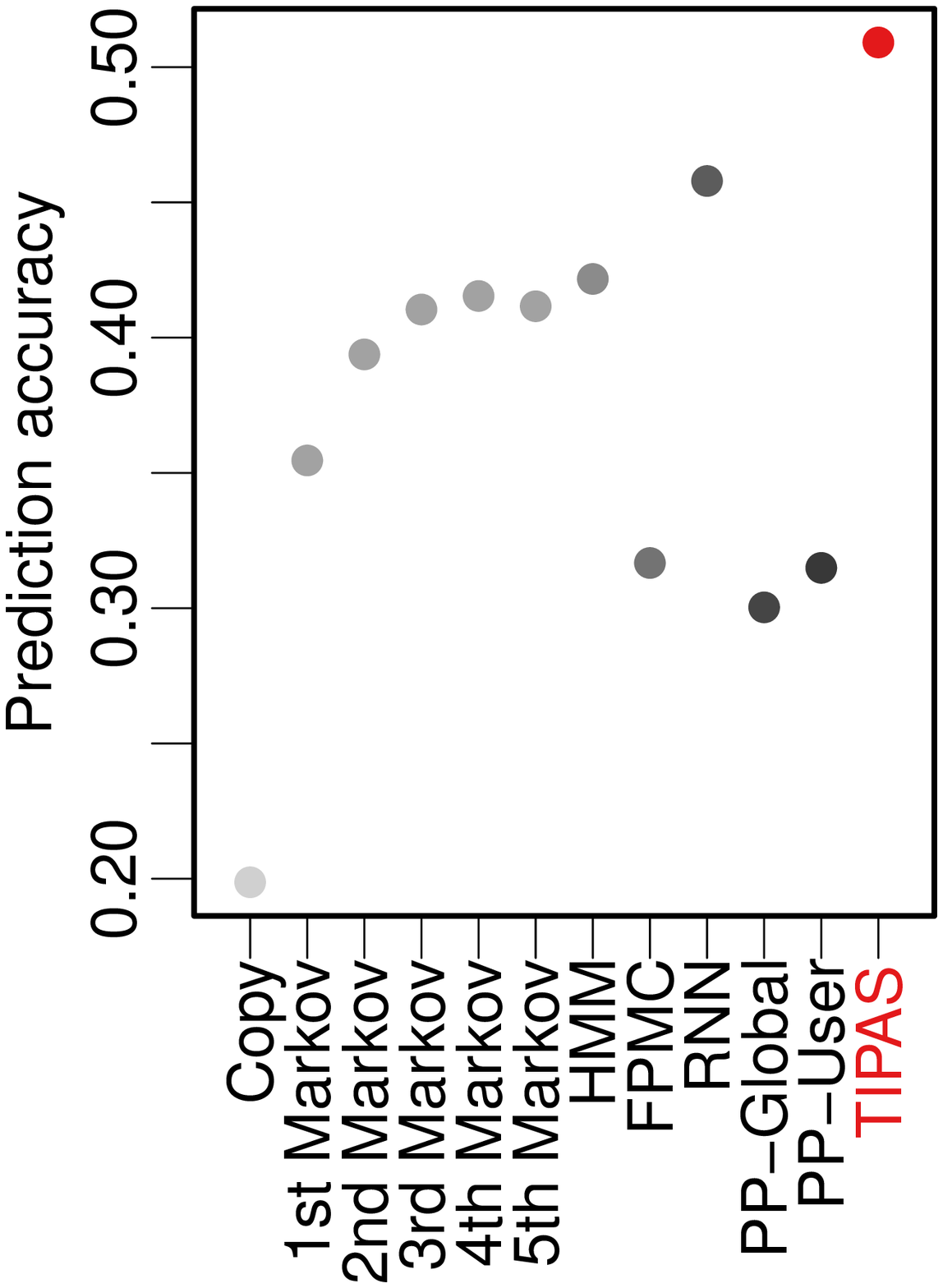}} 
  \vspace{-2mm}
  \caption{Accuracy when predicting actions. Higher is better.
  Comparing proposed TIPAS model (red) to baselines (gray).
   Error bars in all plots correspond to standard errors.
  }
   \vspace{-2mm}
  \label{fig:action_prediction_micro_baselines}
\end{figure}

\xhdr{Results: Comparison to baseline models}
Figure~\ref{fig:action_prediction_micro_baselines} compares accuracy of next action prediction.
We observe that the eleven baselines achieve accuracies of 36-57\% on the Argus dataset and 20-46\% on the Under Armour dataset with the RNN baseline performing best in both datasets.
The limited predictive performance of these competitive baselines shows that this prediction task is non-trivial. 
TIPAS outperforms all baselines on both Argus (60.9\%; 6-69\% rel. improvement) and Under Armour datasets (50.9\%; 11-156\% rel. improvement).
The small standard error across multiple dataset splits in the rolling window evaluation (Figure~\ref{fig:action_prediction_micro_baselines}) demonstrates that
our training procedure is robust and consistently shows good performance. 
We note that TIPAS performs particularly well on rare actions leading to 9-256\% relative improvement in macro-averaged recall over baseline models (Online Appendix~\cite{kurashima2017onlineappendix}).

\begin{figure}[tbp]
  \centering
  \vspace{-1mm}
    \subfigure[Model components (Argus)]{\label{fig:action_prediction_micro_components_a}\centering\includegraphics[width=.42\columnwidth]{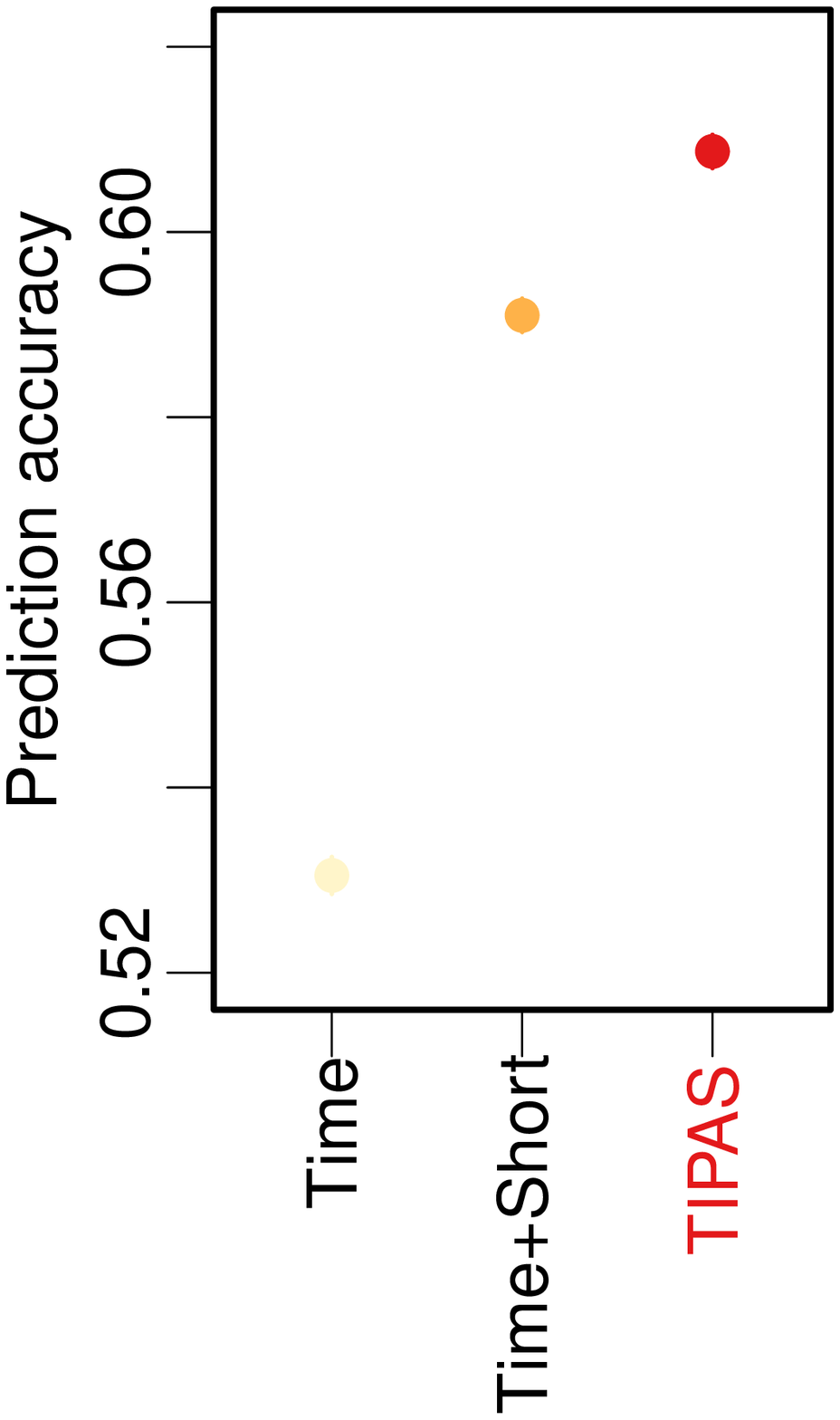}} 
    \hspace{5mm}
  \subfigure[Model components (UA)]{\label{fig:action_prediction_micro_baselines_components_b}\centering\includegraphics[width=.42\columnwidth]{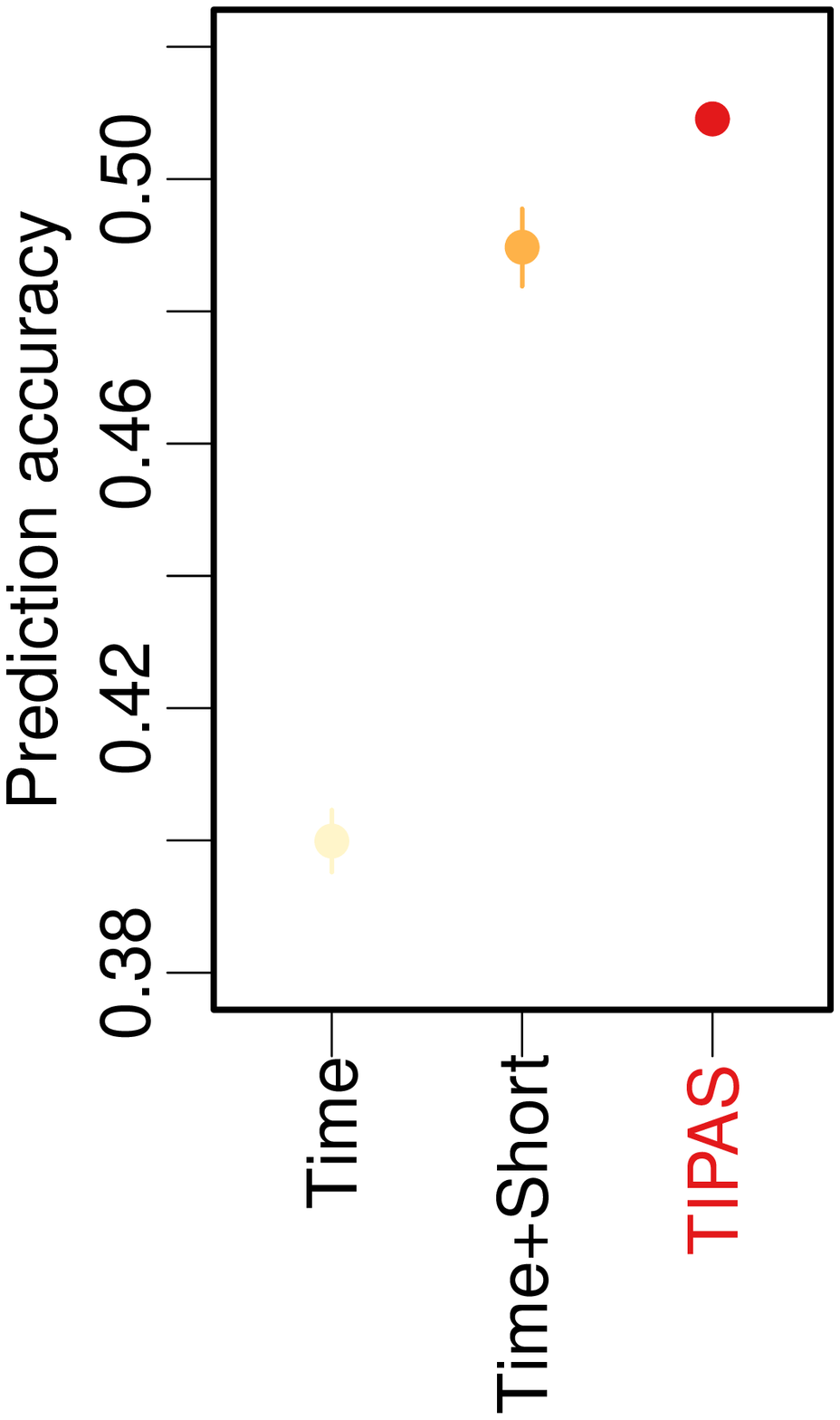}} 
  \vspace{-3mm}
  \caption{
  Ablation study comparing different model components on accuracy when predicting actions. Higher is better.
  }
   \vspace{-2mm}
  \label{fig:action_prediction_micro_components}
\end{figure}

\xhdr{Results: Comparison of individual model components} 
Note that TIPAS has three components (Equation~\ref{eq:lambda}): 
time-varying action propensities (Time), short-term interdependencies between actions (Short), and long-term periodic effects (Long).
Here, we evaluate the performance of each of these components in an ablation study by comparing Time, Time+Short, and the full TIPAS model combining Time+Short+Long (Figure~\ref{fig:action_prediction_micro_components}; all models include user personalized preferences $\alpha_{ua}$). 
We find that modeling time-varying action propensities achieves an accuracy of 53\% and 40\% on the two datasets, respectively.
Further, modeling short-term dependencies between actions improves this to 59\% and 49\%,
and capturing long-term periodicities of actions further improves this to 61\% and 51\%, respectively.
This demonstrates that capturing all three properties is essential to predicting actions in both datasets of human real-world action sequences.
Further, we observe a bigger difference between the full Time+Short+Long model and the Time+Short model in terms of macro-averaged recall (7\% and 5\% relative MAR improvements compared to 3\% and 4\% in terms of accuracy on the Argus and Under Armour datasets, respectively).
This indicates that modeling long-term periodicities is especially important for more rare actions such as walking and biking.
In addition, we find that modeling long-term periodic effects discretized by time of day (0-6h, 6-12h, 12-18h, 18-24h) performs significantly better than not discretizing by time of day on both datasets.
For example, actions such as biking and walking are periodic but vary based on time of day (Figure~\ref{fig:fraction_rayleigh}).
Our full model captures these time-of-day dependent long-term effects and relatively improves macro-averaged recall of predicting biking and walking actions by 491-556\% over Time model and 2-4\% over Time+Short model.

\subsection{Predicting the Time of the Next Action}\label{sec:timeprediction}

We now focus on the second aspect of modeling real-world actions: Predicting the time of the next action.
Specifically, the task is the predict the $n\!+\!1$-th timestamp $t_{u n+1}$ in history $u$,
given past events $H_{u}=\{(a_{u1},t_{u1}), ...,(a_{un},t_{un})\}$ (we do not assume that the next action $a_{u n+1}$ is given).
Mean absolute error (MAE) is used as the evaluation metric.
We use the same train/test paradigm as before (rolling window evaluation training one month and testing on the next).
We restrict predictions to only events that will occur within the next 12 hours (\ie, the time interval $t_{un+1} - t_{un} \leq $ 12 hours) because these are the most important and actionable inferences 
(\eg, predicting a sleep time many days from now may have large error, but it is also less relevant).
In order to make time predictions based on TIPAS, we simulate the multivariate temporal point process using Ogata's modified thinning algorithm~\cite{ogata1981lewis}. 
We simulate 100 samples and return the average time.

We compare our model to the following five baseline methods:
\begin{itemize}
\item \textbf{Time Copy Model:} Predicts the next time, $t_{un+1}$, based on the most recent time-interval of user $u$ $(t_{un+1} = t_{un} + (t_{un} - t_{un-1})).$
\item \textbf{Average Time Interval:} Predicts the next time $t_{un+1}$ using the global average of time-intervals.
\item \textbf{User Average Time Interval:} Predicts the next time $t_{un+1}$ using the average of time-intervals for user $u$.
\item \textbf{PP-Global:} A global Poisson process model. The intensity function is constant over time and defined by $\lambda_{u}(t,a) = \alpha_{a}$. 
\item \textbf{PP-User:} A user-specific Poisson process model. The intensity function is constant over time: $\lambda_{u}(t,a) = \alpha_{ua}$. 
\end{itemize}

We note that the other baselines (Markov models, HMM, FPMC, and RNN) used in Section~\ref{sec:behaviorprediction} are unable to make any time predictions.

\begin{figure}[tbp]
  \centering
  \subfigure[Baseline Comparison (Argus)]{\label{fig:mae_time_a}\centering\includegraphics[width=.44\columnwidth]{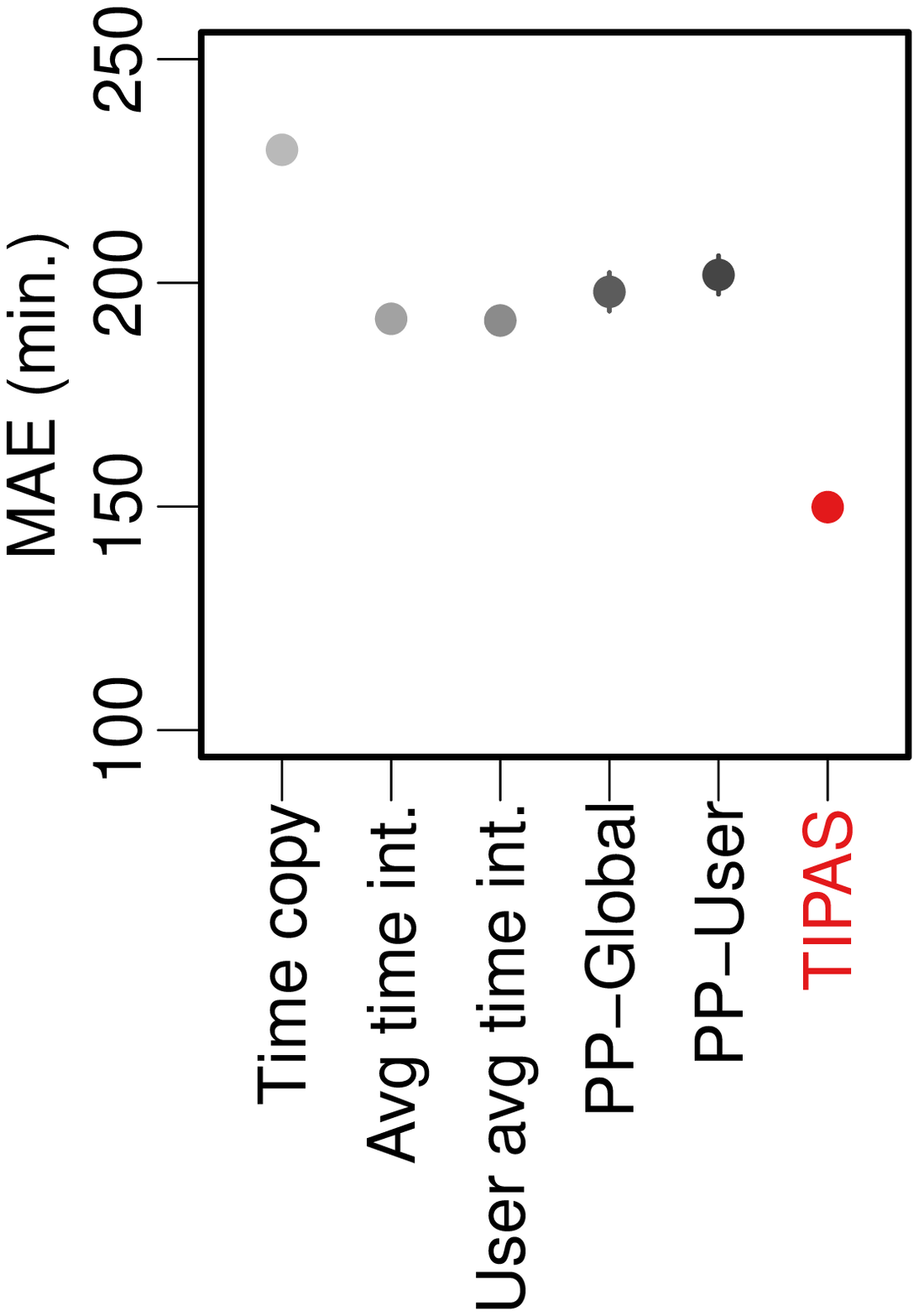}} 
  \hspace{5mm}
  \subfigure[Baseline Comparison (UA)]{\label{fig:mae_time_b}\centering\includegraphics[width=.44\columnwidth]{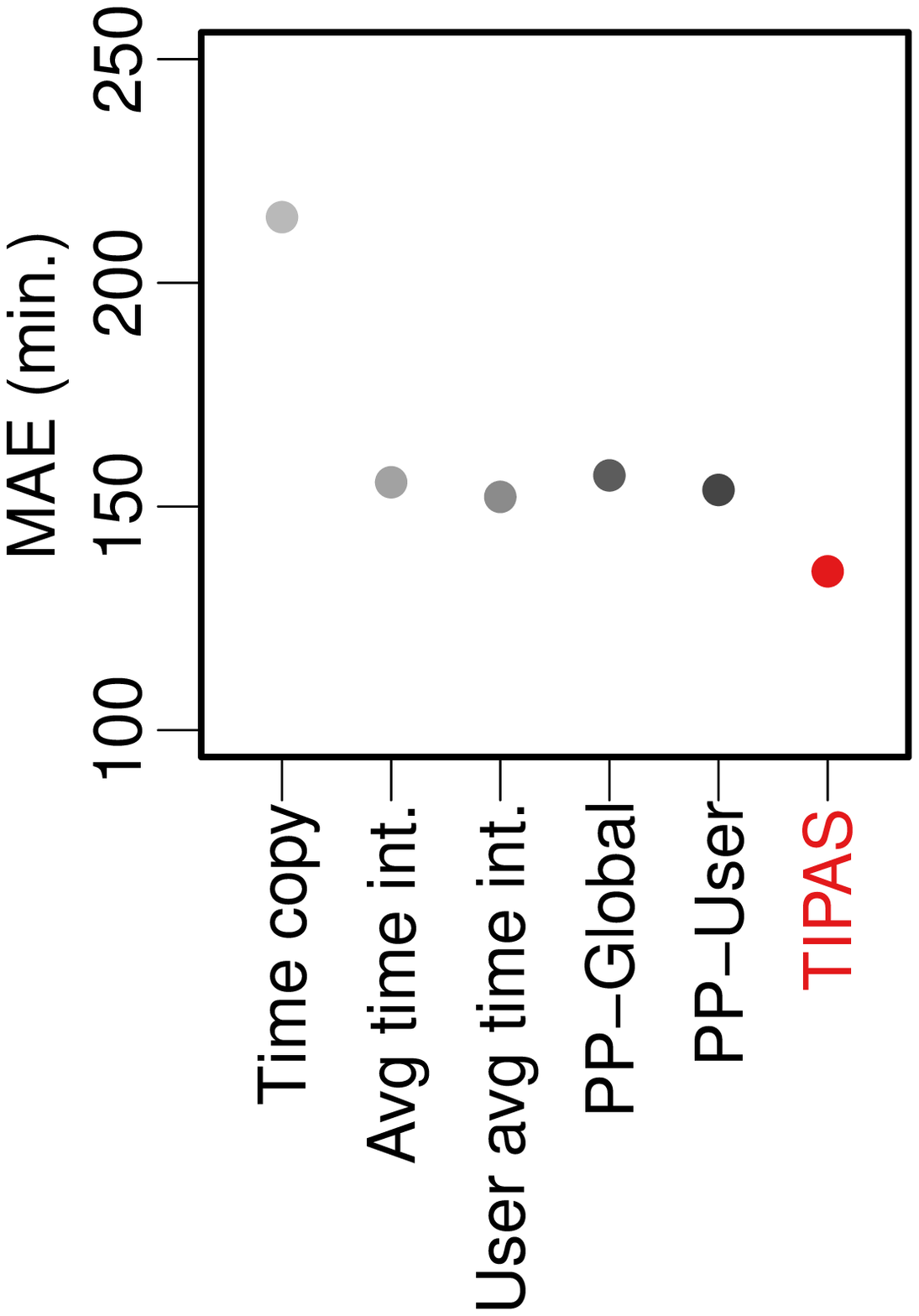}} 
  \vspace{-2mm}
  \caption{Mean absolute error (MAE) when predicting time of next actions. Lower is better. 
  Comparison to baselines. 
  }
  \label{fig:mae_time}
\end{figure}

\xhdr{Results}
Experimental results are shown in Figure~\ref{fig:mae_time}. 
We observe that all baselines perform similarly except the Time Copy model which performs significantly worse on both datasets. 
TIPAS significantly outperforms all baselines across both datasets by 22-35\% in the Argus dataset and 11-37\% in the Under Armour dataset (relative improvement).
Restricting predictions to events within the next 6 hours (instead of 12h as before), TIPAS outperforms the baselines even more significantly, improving upon them by 44-58\% and 37-41\% on the two datasets.
TIPAS is able to make better timing predictions because it is able to leverage three key components.
First, it is aware that certain actions only happen during certain parts of the day. For example, it will predict longer delays in the middle of the night when actions are unlikely to occur.
Second, the model can exploit dependencies between actions. For instance, it might predict a very short time after a run because many users will drink water or check their heart rate soon after.
Third, TIPAS is able to exploit periodicities in the data. For example, it might predict an evening time commute because it observed a commute in the morning. 
In summary, modeling these three key aspects of human behavior allows us to make better predictions of actions and their timing.

\begin{figure}[tbp]
  \centering
  \vspace{-6mm}
  \subfigure[Inferred food periodicity]{\label{fig:interpretable_food_food}\centering\includegraphics[width=.48\columnwidth]{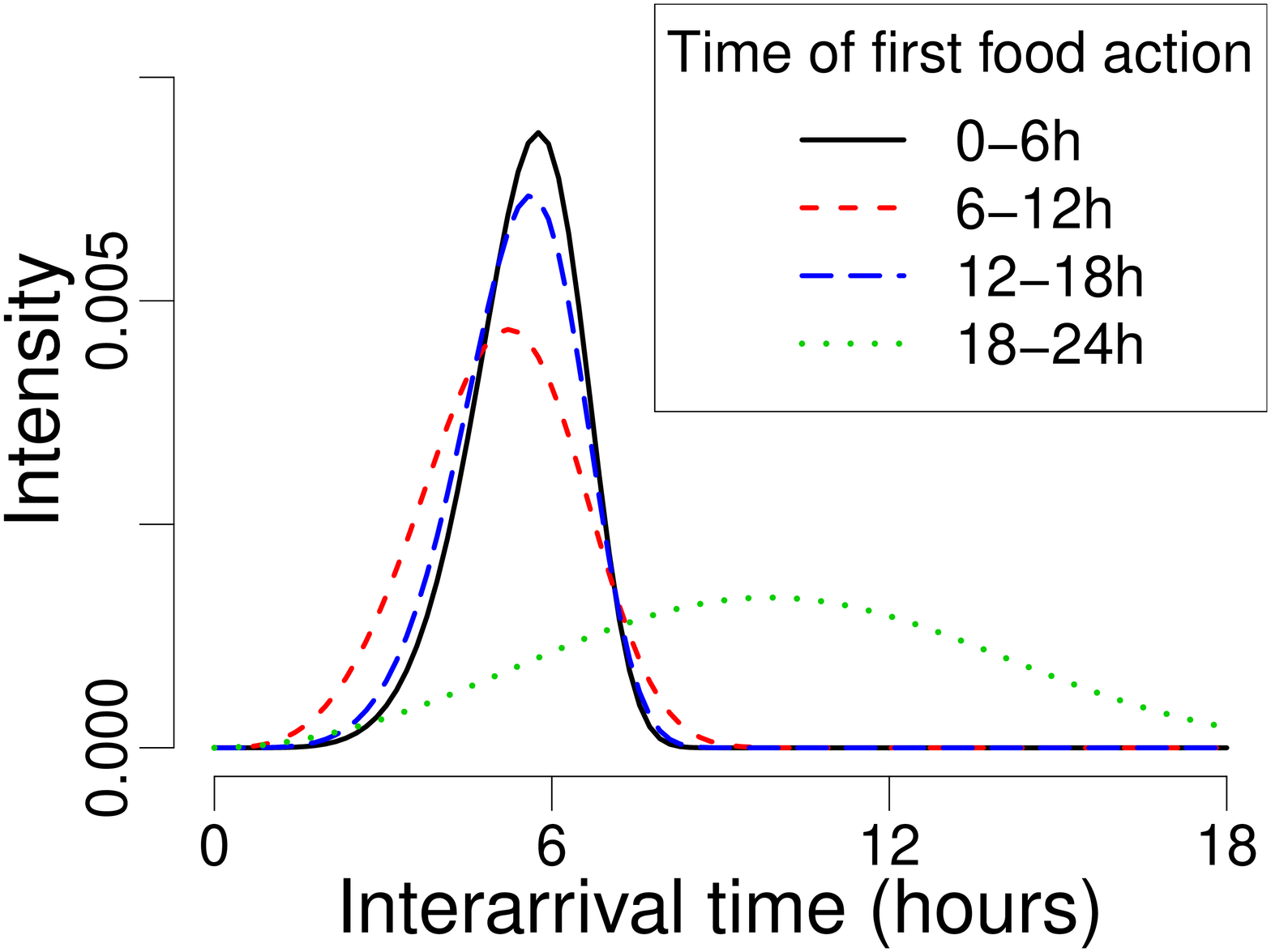}}
  \hspace{1mm}
    \subfigure[Interdependent actions]{\label{fig:interpretable_after_drink}\centering\includegraphics[width=.48\columnwidth]{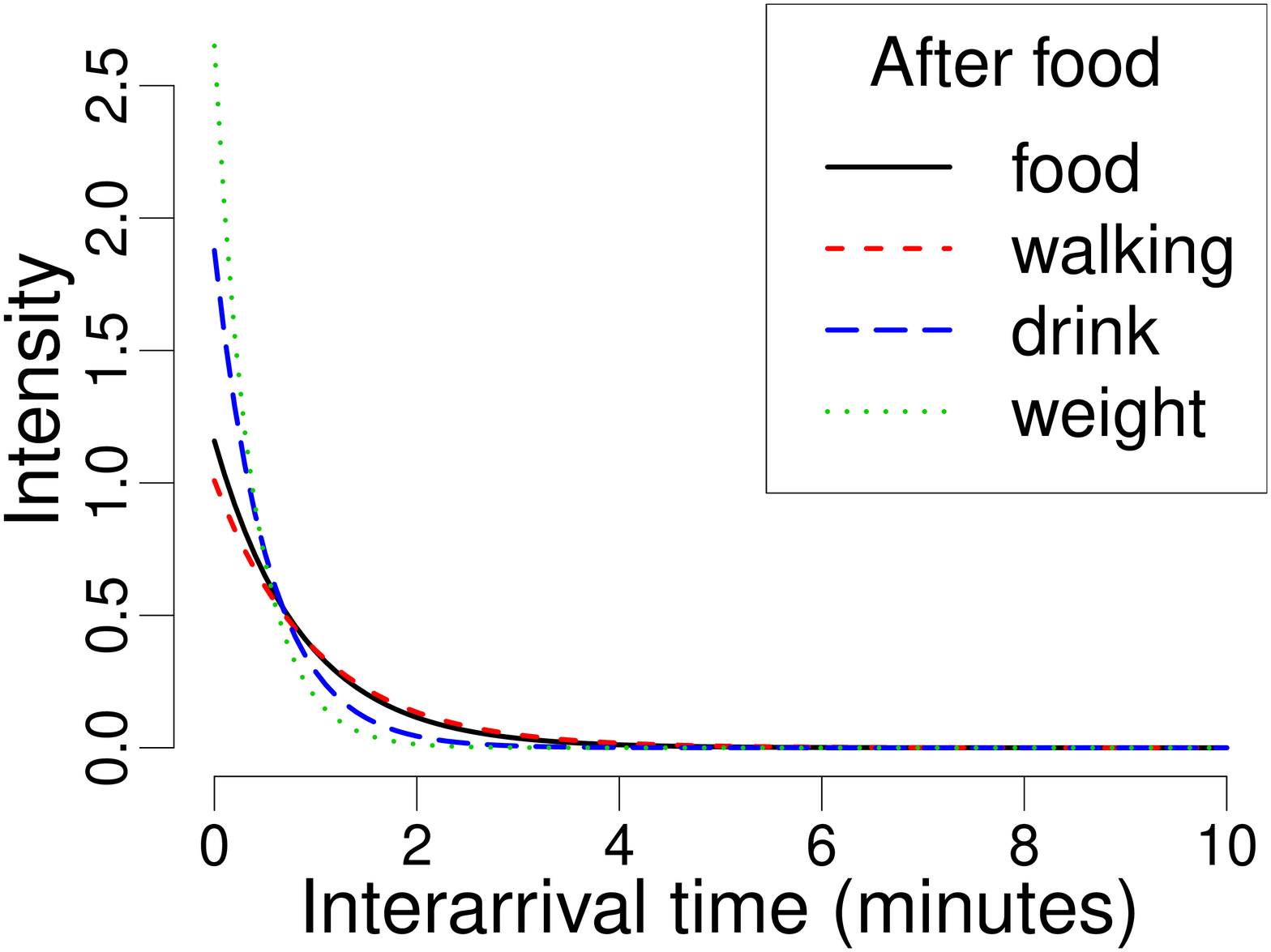}}
  \vspace{-3mm}
  \caption{
  Visualization of inferred TIPAS model parameters for (a) periodicity of food actions and (b) interdependent actions following food actions.
  The learned dependencies allow to explain why specific actions are being predicted.  
  \label{fig:interpretability_examples}
  }
   \vspace{-3mm}
\end{figure}

\subsection{Model Explainability}\label{subsec:interpretability}
TIPAS also allows for visualization of model parameters, which enables explanations of why certain predictions are made. 
This is especially important in the mobile health context, where model predictions may impact users' real-world health behaviors and therefore need to be explained and monitored. 

The inferred model parameters for Equation~\ref{eq:longterm} are shown in Figure~\ref{fig:interpretability_examples}a (specifically, $f(\Delta_{t't})=\gamma_{c_{t'}a} \kappa_{c_{t'}a} \Delta_{t't}^{\kappa_{c_{t'}a}-1} \! \exp(- \gamma_{c_{t'}a} \Delta_{t't}^{\kappa_{c_{t'}a}})$ for $a=\text{food}$).
These distributions correspond to when food events likely trigger other food events.
The distributions show that meals are extremely periodic and that meals sharply determine the timing of the next meal, except for dinners after 18:00h, which do not precisely determine the timing of the next meal (18-24h, green).
The periodicities vary between 5h after breakfast (6-12h) and 6h after lunch (12-18h). This is consistent with a typical schedule of meals at 7:00h, 12:00h, and 18:00h. 
Importantly, this enables us to correctly predict that earlier lunches may lead to earlier dinners.
Such predictions are critical for correctly timed interventions, for instance making sure that diet reminders do not come to late.

Furthermore, TIPAS allows us to explain why an activity was predicted, based on the relative contributions of model components to the overall intensity function (Section~\ref{sec:model_def}).
For example, after food actions users are likely to log other foods and drinks (Figure~\ref{fig:interpretability_examples}b; showing $f(\Delta_{t't})=\omega_{a'a}\exp(-\omega_{a'a}\Delta_{t't})$ for $a'=\text{food}$). 
This makes sense as typical meals include both food and drinks, and users may choose to log parts of each meal separately.
Interestingly, users also often log their weight right after food, indicating that they might be conscious of how their meal might have impacted their weight.
Lastly, we observe walking actions right after meals. 
Users may walk back from a restaurant, or they might attempt to walk off some of their meal's calories. 

While these results and examples are specific to mobile activity logging applications, the utility of our model may generalize other domains where behaviors are time-varying, interdependent, or periodic. 
Distributional choices for the individual may vary across domains but can easily be adapted in our model.

\section{Conclusion}
\label{sec:conclusion}

Accurately predicting the user's future actions is essential for personalization, user modeling, and timely interventions in mobile health applications.
In this paper we demonstrated that real-world user behavior exhibits several complexities including a large number of potential actions, time-varying action propensities, dependencies between actions, and periodic behaviors.
We proposed a novel statistical model based on multivariate temporal point processes that jointly models all these complexities of human behaviors.
Empirically, we demonstrate that our model successfully captures these dynamics in two real-world datasets 
and that it significantly outperforms nine baselines on tasks of predicting the next user action and when this action will occur. 
Our model can serve as a foundation to predict more fine-gained attributes of real-world actions such as their duration, intensity, or exact location.
Our results further have implications for modeling human behavior, app personalization, and targeting of health interventions.

\xhdr{Acknowledgments}
This research has been supported in part by 
NIH BD2K, DARPA NGS2, ARO MURI, IARPA HFC,
Stanford Data Science Initiative, and 
Chan Zuckerberg Biohub.

\clearpage
\pagebreak
\balance

\bibliographystyle{ACM-Reference-Format}
\bibliography{refs} 

\clearpage
\pagebreak
\section{Appendix}
\label{sec:appendix}

\subsection{Details on Model Inference}

We use maximum likelihood estimation to infer the parameters of our proposed model (Equation~\ref{eq:lambda}).
The unknown parameters are
$\bm{\alpha} = \{\{\alpha_{ua}\}_{u \in U}\}_{a \in A}$,
$\bm{\beta} = \{\{\beta_{az}\}_{a \in A}\}_{z \in \bm{Z}}$,
$\bm{\mu} = \{\{\mu_{az}\}_{a \in A}\}_{z \in \bm{Z}}$,
$\bm{\sigma} = \{\{\sigma_{az}\}_{a \in A}\}_{z \in \bm{Z}}$,
$\bm{\Theta} = \{\{\theta_{a'a}\}_{a \in A}\}_{a' \in A}$,
$\bm{\Omega} = \{\{\omega_{a'a}\}_{a \in A}\}_{a' \in A}$,
$\bm{\Phi} = \{\{\phi_{ca}\}_{c \in C}\}_{a \in A}$,
$\bm{\Gamma} = \{\{\gamma_{ca}\}_{c \in C}\}_{a \in A}$,
and $\bm{K} = \{\{\kappa_{ca}\}_{c \in C}\}_{a \in A}$.
The set of all parameters is denoted by $\bm{\Psi}=\{\bm{\alpha}, \bm{\beta}, \bm{\mu}, \bm{\sigma}, \bm{\Theta}, \bm{\Omega}, \bm{\Phi}, \bm{\Gamma}, \bm{K}\}$.

The log-likelihood function (Equation~\ref{eq:general_log_likelihood}), given a set of user histories $\mathcal{H} = \{H_u\}_{u \in U}$, can be expressed as:
{\small
\vspace{-.25\baselineskip}
\begin{align}
  \mathcal{L}(\bm{\Psi}|\mathcal{H}) = \sum_{u \in U}\sum_{n=1}^{N_{u}} \log \lambda_{u}(t_{un},a_{un}) - \sum_{u \in U} \int_{0}^{T} \sum_{a \in A} \lambda_{u}(t,a)dt \;\;,
  \label{eq:loglikelihood1_appendix}
\end{align}
}
where the last term, the expectation function, represents the expected number of events in the time period from 0 to $T$.
Combining Equations (\ref{eq:lambda})-(\ref{eq:longterm}) and (\ref{eq:loglikelihood1_appendix}),
the log-likelihood can be written as follows:

{\small
\vspace{-.25\baselineskip}
\begin{align}
  \lefteqn{\mathcal{L}(\bm{\Psi}|\mathcal{H}) = }  \nonumber \\
  & \sum_{u \in U} \sum_{n=1}^{N_{u}} \log \Biggl\{ \alpha_{ua_{un}} + \sum_{z \in \bm{Z}} \frac{\beta_{a_{un}z}}{\sqrt{2\pi\sigma_{a_{un}z}^{2}}}\exp\Bigl( - \frac{\bigl(l_{t_{un}}-\mu_{a_{un}z}\bigr)^{2}}{2\sigma_{a_{un}z}^{2}}\
\Bigr) \nonumber \\
  & + \sum_{m=1}^{n-1} \theta_{a_{um}a_{un}} \omega_{a_{um}a_{un}} \exp (-\omega_{a_{um}a_{un}} \Delta_{t_{um}t_{un}})\nonumber \\
  & + \sum_{l=1}^{n-1} \Bigl( I(a_{ul}=a_{un}) \phi_{c_{ul}a_{un}} \gamma_{c_{ul}a_{un}} \kappa_{c_{ul}a_{un}} \Delta_{t_{ul}t_{un}}^{\kappa_{c_{ul}a_{un}}-1}  \nonumber \\
  & \times \exp( -\gamma_{c_{ul}a_{un}} \Delta_{t_{ul}t_{un}}^{\kappa_{c_{ul}a_{un}}} ) \Bigr) \Biggr\} - \sum_{u \in U} \int_{0}^{T} \sum_{a \in A} \lambda_{u}(t,a)dt \;\;,
  \label{eq:loglikelihood2_appendix}
\end{align}
}
where $c_{ul} \in C$ represents time-of-day category of $l$-th event of $u$,
and $I(\cdot)$ is the indicator function. 
We can analytically calculate the integral in Equation~\ref{eq:loglikelihood2_appendix} as follows:
{\small
\vspace{-.25\baselineskip}
\begin{align}
  \lefteqn{ \sum_{u \in U} \int_{0}^{T} \sum_{a \in A} \lambda_{u}(t,a) dt = T \sum_{u \in U} \sum_{a \in A} \alpha_{ua} } \nonumber \\
  & + |U|\frac{T}{\mathcal{T}}\sum_{a \in A}\sum_{z \in \bm{Z}} \frac{\beta_{az}}{2} \bigl( \mathit{erf}(\frac{\mu_{az}}{\sqrt{2}\sigma_{az}}) + \mathit{erf}(\frac{\mathcal{T} - \mu_{az}}{\sqrt{2}\sigma_{az}}) \bigr) \nonumber \\
  & + \sum_{u \in U} \sum_{a \in A} \sum_{n=1}^{N_{u}} \theta_{a_{un}a} \Bigl( 1 - \exp\bigl(\omega_{a_{un}a}(T-t_{un}\bigr)\Bigr) \nonumber \\
  & + \sum_{u \in U} \sum_{n=1}^{N_{u}} \phi_{c_{un}a_{un}} \Bigl( 1 -\exp\bigl(- \gamma_{c_{un}a_{un}} (T-t_{un})^{\kappa_{c_{un}a_{un}}} \bigr)\Bigr) \;\;,
  \label{eq:exponential_in_expectation_function}
\end{align}
}
where $\mathcal{T}$ is the time period of a day (\ie, 24 hours), 
$\frac{T}{\mathcal{T}}$ is the number of days representation of the observed period $T$,
and where $\mathit{erf}$ denotes the Gauss error function $\mathit{erf}(x) = \frac{1}{\sqrt\pi}\int_{-x}^x e^{-t^2} \,\mathrm dt $.

Inspired by previous work~\cite{zhou2013learning,farajtabar2015coevolve}, we develop an efficient inference algorithm to maximize the log-likelihood based on the EM algorithm.
By iterating the E-step and the M-step until convergence,
we obtain a local optimum solution for $\bm{\Psi}$.

\xhdr{E-step}
Conceptually, we introduce latent variables $\bm{p},\bm{q},\bm{r}$ to capture why each event was triggered either through user preference, time-varying background intensity, short-term action interdependencies, or long-term periodic effects.
Let $p_{0,un}$ be the probability that the $n$-th event of user $u$ was triggered by user preference,
$p_{z,un}$ be the probability that the $n$-th event of user $u$ was triggered by the time-varying background intensity function of latent class $z$,
$q_{um,un}$ be the probability that the $n$-th event of user $u$ was triggered by the short-term effect of the $m$-th event of user $u$,
and $r_{ul,un}$ be the probability that the $n$-th event of user $u$ was triggered by the long-term effect of the $l$-th event of user $u$.

In E-step, $k$-th estimate of $p^{k}_{0,un}$, $p_{z,un}^{k}$, $q^{k}_{um,un}$, and $r^{k}_{ul,un}$ are calculated by:
{\small
\vspace{-.25\baselineskip}
\begin{align}
  p^{k}_{0,un} = \frac{\alpha^{k}_{ua_{un}}}{R_{un}} \;\;,
\end{align}
\vspace{-.75\baselineskip}
\begin{align}
  p_{z,un}^{k} = \frac{1}{R_{un}} \frac{\beta_{a_{un}z}^{k}}{\sqrt{2\pi(\sigma_{a_{un}z}^{k})^{2}}}\exp\Bigl( - \frac{\bigl(l_{t_{un}}-\mu^{k}_{a_{un}z}\bigr)^{2}}{2(\sigma^{k}_{a_{un}z})^{2}}\Bigr) \;\;,
\end{align}
\vspace{-.75\baselineskip}
\begin{align}
  q^{k}_{um,un} = \frac{1}{R_{un}} \theta^{k}_{a_{um}a_{un}} \omega^{k}_{a_{um}a_{un}} \exp (-\omega^{k}_{a_{um}a_{un}} \Delta_{t_{um}t_{un}}) \;\;,
\end{align}
\vspace{-.75\baselineskip}
\begin{align}
  \lefteqn{ r^{k}_{ul,un} = \Biggl\{ \frac{1}{R_{un}} \phi^{k}_{c_{ul}a_{un}} \gamma^{k}_{c_{ul}a_{un}} \kappa^{k}_{c_{ul}a_{un}} } \nonumber \\
  & \quad \qquad \qquad \times \Delta_{t_{ul}t_{un}}^{\kappa^{k}_{c_{ul}a_{un}} - 1} \exp(- \gamma^{k}_{c_{ul}a_{un}} \Delta_{t_{ul}t_{un}}^{\kappa^{k}_{c_{ul}a_{un}}}) \Biggr\}\;\;,
\end{align}
}
where $\bm{\Psi}^{k}=\{\bm{\alpha}^{k}, \bm{\beta}^{k}, \bm{\mu}^{k}, \bm{\sigma}^{k}, \bm{\Theta}^{k}, \bm{\Omega}^{k}, \bm{\Phi}^{k}, \bm{\Gamma}^{k}, \bm{K}^{k}\}$ is the $k$-th estimate of parameters in the EM procedure,
and $R_{un}$ is the normalization factor in order to satisfy $p^{k}_{0,un} + \sum_{z \in \bm{Z}} p^{k}_{z,un} + \sum_{m=1}^{n-1} q^{k}_{um,un} + \sum_{l=1}^{n-1} r^{k}_{ul,un} = 1$.

\xhdr{M-step}
We use Jensen's inequality to maximize a lower bound on the log-likelihood (Equation~\ref{eq:loglikelihood2_appendix}) in the M-step which is as follows:
{\small
\vspace{-.25\baselineskip}
\begin{align}
  \lefteqn{ Q(\bm{\Psi}|\bm{\Psi}^{k}) =} \nonumber \\
  & \sum_{u \in U} \sum_{n=1}^{N_{u}} \Biggl\{ p^{k}_{0,un} \log \frac{\alpha_{ua_{un}}}{p^{k}_{0,un}}
  + \sum_{z \in \bm{Z}} p^{k}_{z,un} \log \frac{f_{z}(t_{un},a_{un}))}{p^{k}_{z,un}} \nonumber \\
  & + \sum_{m=1}^{n-1} q^{k}_{um,un} \log \frac{ g_{a_{um}a_{un}}(\Delta_{t_{um}t_{un}})}{q^{k}_{um,un}} \nonumber \\
  & + \sum_{l=1}^{n-1} I(a_{ul}=a_{un}) r^{k}_{ul,un} \log \frac{h_{t_{ul}a_{un}}(\Delta_{t_{ul}t_{un}})}{r^{k}_{ul,un}} \Biggr\} \nonumber \\
  & - \sum_{u \in U} \int_{0}^{T} \sum_{a \in A} \lambda_{u}(t,a)dt \;\;,
  \label{eq:qfunction}
\end{align}
}
where $f,g,h$ represent the unnormalized intensity that the $n$-th event of user $u$ was triggered by the time-varying action propensity, short-term dependencies, or long-term periodic effects, respectively; they are defined as follows:

\begin{eqnarray}
  f_{z}(t,a) = \frac{\beta_{az}}{\sqrt{2\pi\sigma_{az}^{2}}}\exp\Bigl( - \frac{\bigl(l_{t}-\mu_{az}\bigr)^{2}}{2\sigma_{az}^{2}}\Bigr)\;\;,
\end{eqnarray}
\begin{align}
  g_{a'a}(\Delta_{t't}) = \theta_{a'a}\omega_{a'a}\exp(-\omega_{a'a}\Delta_{t't}) \;\;,
\end{align}
\begin{align}
  h_{t'a}(\Delta_{t't}) = \phi_{c_{t'}a} \gamma_{c_{t'}a} \kappa_{c_{t'}a} \Delta_{t't}^{\kappa_{c_{t'}a}-1} \exp(- \gamma_{c_{t'}a} \Delta_{t't}^{\kappa_{c_{t'}a}}) \;\;.
\end{align}

We obtain the next estimate of the parameters by taking the derivative of the lower bound on the log-likelihood with respect to each parameter and setting them to zero.
The lower bounds on the Q function with respect to $\bm{\alpha}$, $\bm{\beta}$, $\bm{\Theta}$ and $\bm{\Phi}$ are as follows.
{\small
  \vspace{-.25\baselineskip}
\begin{align}
  Q(\bm{\alpha}|\bm{\Psi}^{k}) = \sum_{u \in U} \sum_{n=1}^{N_{u}} p^{k}_{0,un} \log \frac{\alpha_{ua_{un}}}{p^{k}_{0,un}} - T \sum_{u \in U} \sum_{a \in A} \alpha_{ua} \;\;,
  \label{eq:qfunction_alpha}
\end{align}
}
{\small
\vspace{-.25\baselineskip}
\begin{align}
  \lefteqn{ Q(\bm{\beta}|\bm{\Psi}^{k}) =
    \sum_{u \in U} \sum_{n=1}^{N_{u}} \sum_{z \in \bm{Z}} p^{k}_{z,un} \log \frac{f_{z}(t_{un},a_{un}))}{p^{k}_{z,un}} } \nonumber \\
  & - |U|\frac{T}{\mathcal{T}}\sum_{a \in A}\sum_{z \in \bm{Z}} \frac{\beta_{az}}{2} \bigl( \mathit{erf}(\frac{\mu_{az}}{\sqrt{2}\sigma_{az}}) + \mathit{erf}(\frac{\mathcal{T} - \mu_{az}}{\sqrt{2}\sigma_{az}}) \bigr) \;\;,
  \label{eq:qfunction_beta}
\end{align}
}
{\small
  \vspace{-.25\baselineskip}
  \begin{align}
    \lefteqn{ Q(\bm{\theta}|\bm{\Psi}^{k}) =
      \sum_{u \in U} \sum_{n=1}^{N_{u}} \sum_{m=1}^{n-1} q^{k}_{um,un} \log \frac{ g_{a_{um}a_{un}}(\Delta_{t_{um}t_{un}})}{q^{k}_{um,un}} } \nonumber \\
    & - \sum_{u \in U} \sum_{a \in A} \sum_{n=1}^{N_{u}} \theta_{a_{un}a} \Bigl( 1 - \exp\bigl(\omega_{a_{un}a}(T-t_{un}\bigr)\Bigr) \;\;,
    \label{eq:qfunction_theta}
  \end{align}
}
{\small
\vspace{-.25\baselineskip}
\begin{align}
  \lefteqn{ Q(\bm{\phi}|\bm{\Psi}^{k}) = 
    \sum_{u \in U} \sum_{n=1}^{N_{u}} \sum_{l=1}^{n-1} I(a_{ul}=a_{un}) r^{k}_{ul,un} \log \frac{h_{t_{ul}a_{un}}(\Delta_{t_{ul}t_{un}})}{r^{k}_{ul,un}} } \nonumber \\
  & - \sum_{u \in U} \sum_{n=1}^{N_{u}} \phi_{c_{un}a_{un}} \Bigl( 1 -\exp\bigl(- \gamma_{c_{un}a_{un}} (T-t_{un})^{\kappa_{c_{un}a_{un}}} \bigr)\Bigr) \;\;.
  \label{eq:qfunction_phi}
\end{align}
}

Their update rules are calculated as follows:
{\small
\vspace{-.25\baselineskip}
\begin{align}
  \alpha_{ua}^{k+1} = \frac{\sum_{n=1}^{N_{u}}I(a_{un}=a)p_{0,un}^{k}}{T} \;\;,
\end{align}
\vspace{-.75\baselineskip}
\begin{align}
  \beta_{az}^{k+1} = \frac{2 \mathcal{T}}{|U| T} \times \frac{\sum_{u \in U} \sum_{n=1}^{N_{u}} I(a_{un}=a) p_{z,un}^{k}}
       { \mathit{erf}(\frac{\mu_{az}^{k}}{\sqrt{2}\sigma_{az}^{k}}) + \mathit{erf}(\frac{\mathcal{T} - \mu_{az}^{k}}{\sqrt{2}\sigma_{az}^{k}}) } \;\;,
\end{align}
\vspace{-.75\baselineskip}
\begin{align}
  \theta_{a'a}^{k+1} =
\frac{\sum_{u \in U} \sum_{n=1}^{N_{u}} \sum_{m=1}^{n-1} I(a_{um}=a',a_{un}=a) q^{k}_{um,un}}
{\sum_{u \in U} \sum_{n=1}^{N_{u}} I(a_{un}=a') \Bigl( 1 - \exp \bigl( - \omega_{a'a}^{k} (T-t_{un}) \bigr) \Bigr)} \;\;,
\end{align}
\vspace{-.75\baselineskip}
\begin{align}
  \phi_{ca}^{k+1} =
  \frac{\sum_{u \in U} \sum_{n=1}^{N_{u}} \sum_{l=1}^{n-1} I(a_{ul}=a,a_{un}=a,c_{ul}=c) r^{k}_{ul,un}}
       {\sum_{u \in U} \sum_{n=1}^{N_{u}} I(a_{un}=a,c_{un}=c) \Bigl( 1 - \exp \bigl( -\gamma_{ca}^{k}(T-t_{un})^{\kappa_{ca}^{k}} \bigr) \Bigr)} \;\;.
\end{align}
}
Because of the exponentials ($\exp$ and $\mathit{erf}$) within the expectation function (Equation~\ref{eq:exponential_in_expectation_function}),
$\omega_{a'a}^{k+1}$, $\gamma_{ca}^{k+1}$, $\kappa_{ca}^{k+1}$, $\mu^{k+1}_{az}$, and $\sigma^{k+1}_{az}$ cannot be solved in closed form.
However, by further considering a lower bound for these exponentials $\omega_{a'a}^{k+1}$ and $\gamma_{ca}^{k+1}$ can be solved in closed form.
The lower bounds on the Q function with respect to $\bm{\Omega}$ and $\bm{\Gamma}$ are as follows.

{\small
  \vspace{-.25\baselineskip}
\begin{align}
  \lefteqn{ Q(\bm{\Omega}|\bm{\Psi}^{k}) =
    \sum_{u \in U} \sum_{n=1}^{N_{u}} \sum_{m=1}^{n-1} q^{k}_{um,un} \log \frac{ g_{a_{um}a_{un}}(\Delta_{t_{um}t_{un}})}{q^{k}_{um,un}} } \nonumber \\
  & - \sum_{u \in U} \sum_{a \in A} \sum_{n=1}^{N_{u}} \theta_{a_{un}a} \omega_{a_{un}a} ( T - t_{un} ) \exp( - \omega^{k}_{a_{un}a} ( T - t_{un} ) ) \;\;,
\label{eq:qfunction_omega}
\end{align}
}
{\small
\vspace{-.25\baselineskip}
\begin{align}
  \lefteqn{ Q(\bm{\Gamma}|\bm{\Psi}^{k}) = } \nonumber\\
  & \sum_{u \in U} \sum_{n=1}^{N_{u}} \sum_{l=1}^{n-1} I(a_{ul}=a_{un}) r^{k}_{ul,un} \log \frac{h_{t_{ul}a_{un}}(\Delta_{t_{ul}t_{un}})}{r^{k}_{ul,un}} \nonumber \\
  & - \sum_{u \in U} \sum_{n=1}^{N_{u}} \Biggl\{ \phi_{c_{un}a_{un}} \gamma_{c_{un}a_{un}} ( T - t_{un} )^{\kappa_{c_{un}a_{un}}} \nonumber \\
  & \times \exp\bigl( - \gamma_{c_{un}a_{un}}^{k} ( T - t_{un} )^{\kappa_{c_{un}a_{un}}} \bigr) \Biggr\} \;\;.
  \label{eq:qfunction_gamma}
\end{align}
}

We obtain the next estimate of $\omega_{a'a}^{k+1}$ and $\gamma_{ca}^{k+1}$  by taking the derivative of these functions and setting them to zero.
Their update rules are calculated as follows:
{\small
\vspace{-.25\baselineskip}
\begin{align}
  \lefteqn{ \omega_{a'a}^{k+1} = \Biggl\{
    \sum_{u \in U} \sum_{n=1}^{N_{u}} \sum_{m=1}^{n-1} I(a_{um}=a',a_{un}=a) q^{k}_{um,un}
    \Biggr\} } \nonumber \\
  & / \Biggl\{ \sum_{u \in U} \sum_{n=1}^{N_{u}} \sum_{m=1}^{n-1} I(a_{um}=a',a_{un}=a) q^{k}_{um,un} \Delta_{t_{um}t_{un}} \nonumber \\
  & + \sum_{u \in U} \sum_{n=1}^{N_{u}} I(a_{un}=a') \theta_{a'a}^{k} (T-t_{un})\exp\bigl(-\omega_{a'a}^{k}(T-t_{un})\bigr) \Biggr\} \;\;,
\end{align}
\begin{align}
  \lefteqn{ \gamma_{ca}^{k+1} = \Biggl\{ \sum_{u \in U} \sum_{n=1}^{N_{u}} \sum_{l=1}^{n-1} I(a_{ul}=a,a_{un}=a,c_{ul}=c) r^{k}_{ul,un} \Biggr\} } \nonumber \\
  & / \Biggl\{ \sum_{u \in U} \sum_{n=1}^{N_{u}} \sum_{l=1}^{n-1} I(a_{ul}=a,a_{un}=a,c_{ul}=c) r^{k}_{ul,un} \Delta_{t_{ul}t_{un}}^{\kappa^{k}_{ca}} \nonumber \\
  & + \sum_{u \in U} \sum_{n=1}^{N_{u}} I(a_{un}=a, c_{un}=c) \phi^{k}_{ca} ( T - t_{un})^{\kappa^{k}_{ca}} \exp\bigl( - \gamma_{ca}^{k} (T-t_{un})^{\kappa^{k}_{ca}}\bigr) \Biggr\} \;\;.
\end{align}
}

The other three parameters, $\kappa_{ca}^{k+1}$, $\mu^{k+1}_{az}$ and $\sigma^{k+1}_{az}$, are estimated by maximizing $Q$ (Equation~\ref{eq:qfunction})
through the use of a gradient-based numerical optimization method; we used the Newton method.

\subsection{Details on Action Prediction and Macro-average Recall}

\begin{figure}[tbp]
  \centering
  \subfigure[Baseline Comparison (Argus)]{\label{fig:macro_avg_recall_baselines_a}\centering\includegraphics[width=.40\columnwidth]{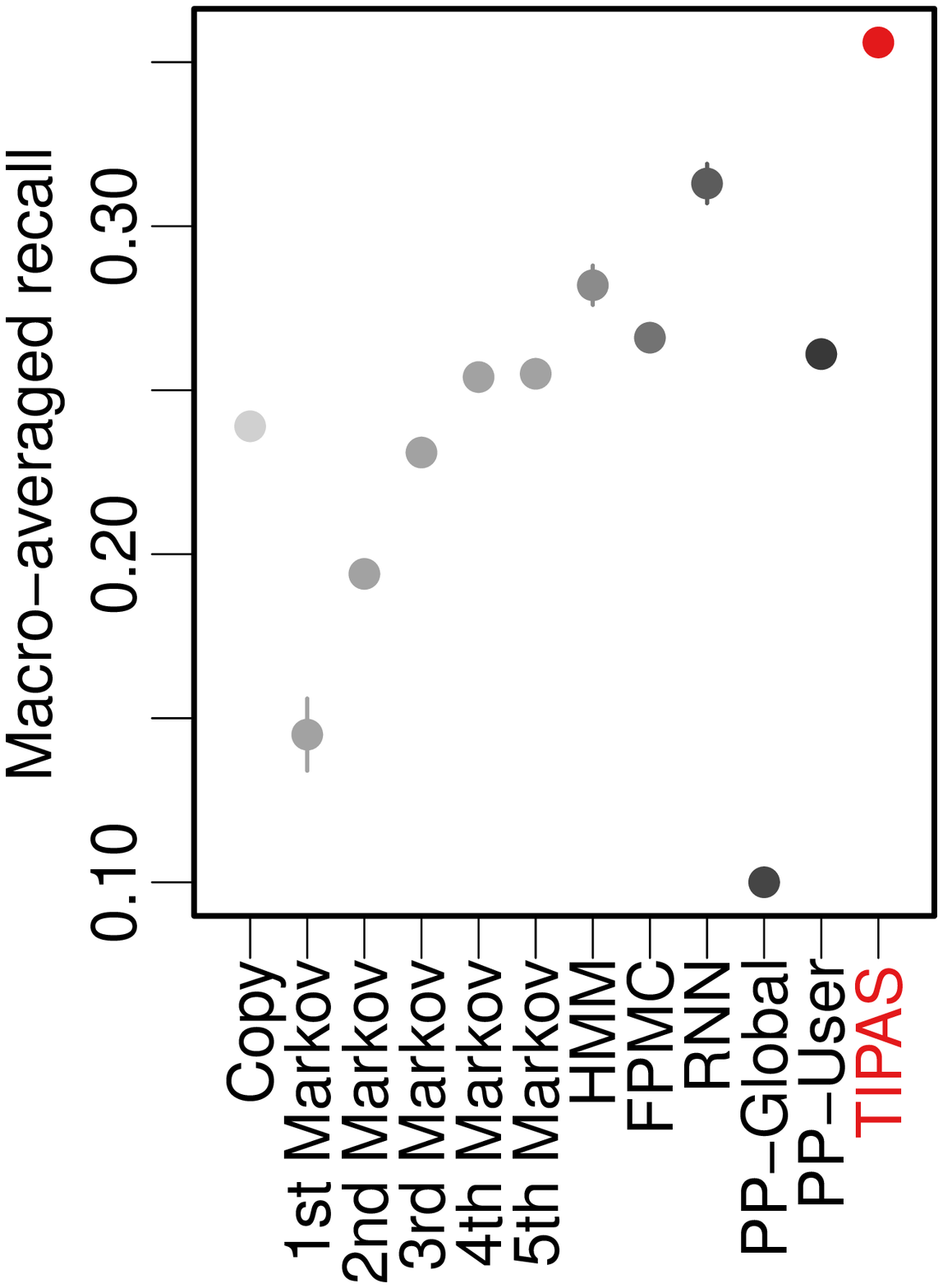}}
    \hspace{5mm}
  \subfigure[Baseline Comparison (UA)]{\label{fig:macro_avg_recall_baselines_b}\centering\includegraphics[width=.40\columnwidth]{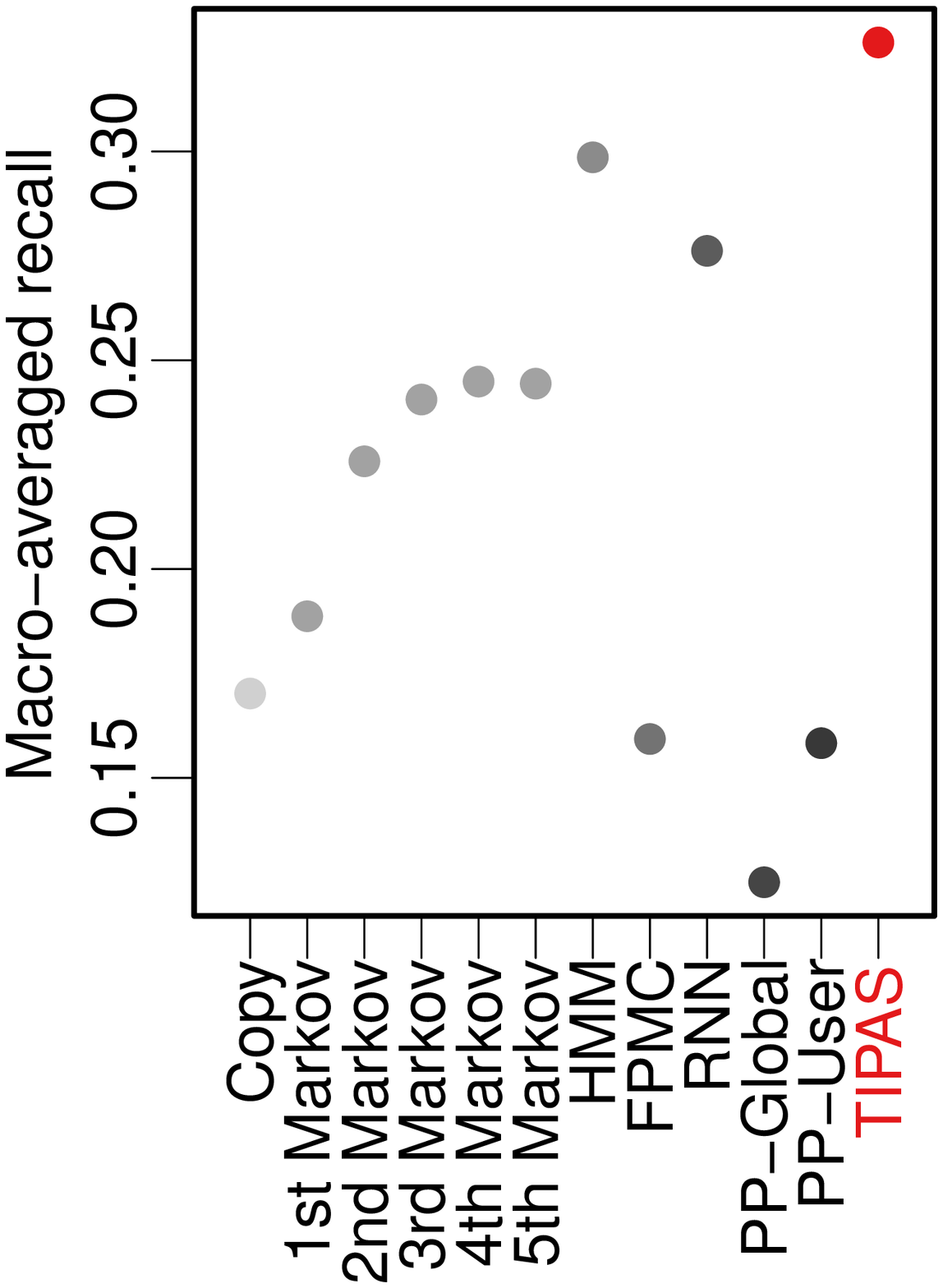}}
  \vspace{-3mm}
  \caption{Macro-averaged recall when predicting action. Higher is better.
  Comparing proposed model TIPAS (red) to baselines (gray).
  Error bars in all plots correspond to standard errors.
  }
  \label{fig:macro_avg_recall_baselines}
\end{figure}

\begin{figure}[tbp]
  \centering
  \subfigure[Model components (Argus)]{\label{fig:macro_avg_recall_components_a}\centering\includegraphics[width=.40\columnwidth]{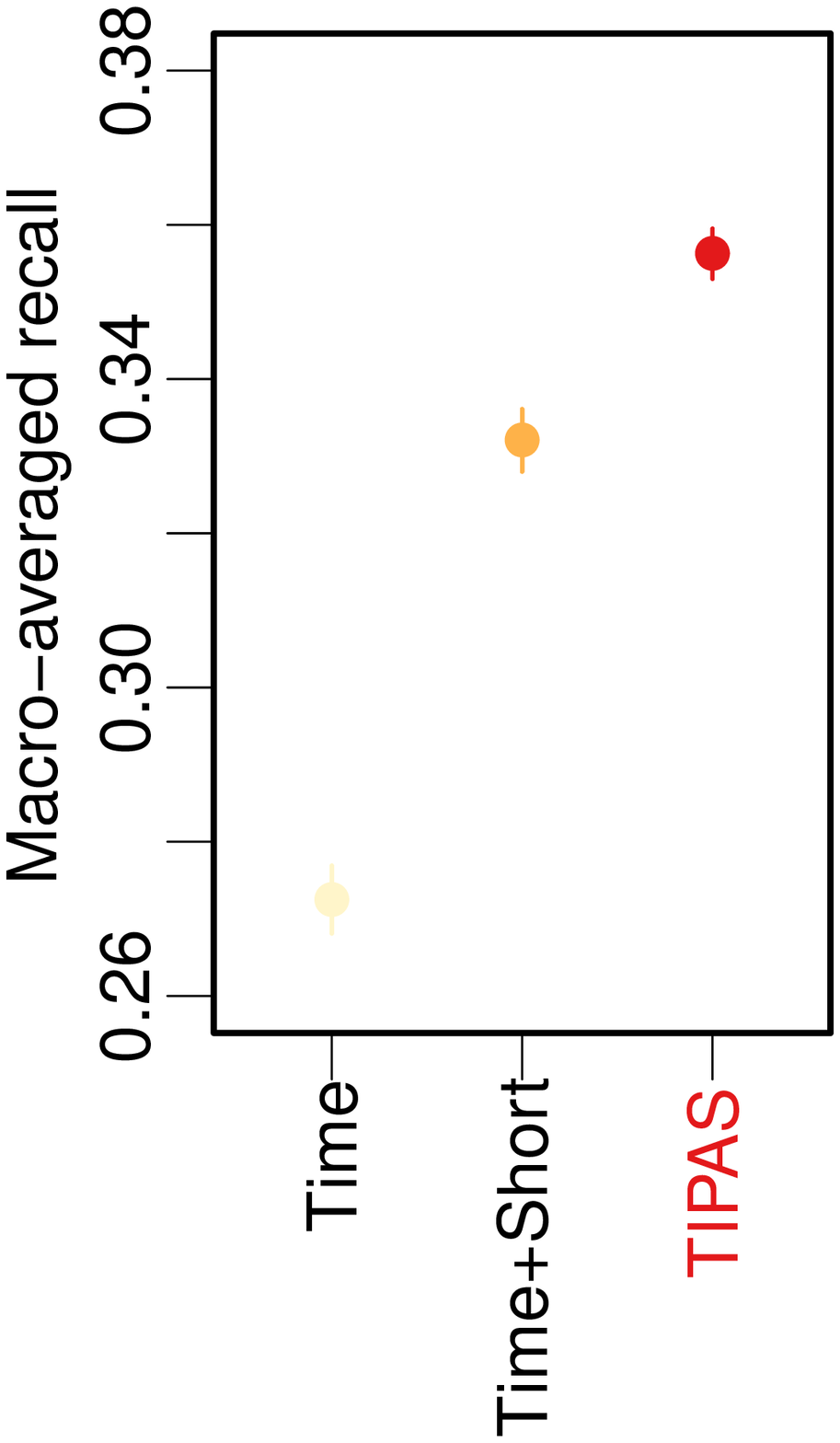}}
  \hspace{5mm}
  \subfigure[Model components (UA)]{\label{fig:macro_avg_recall_components_b}\centering\includegraphics[width=.40\columnwidth]{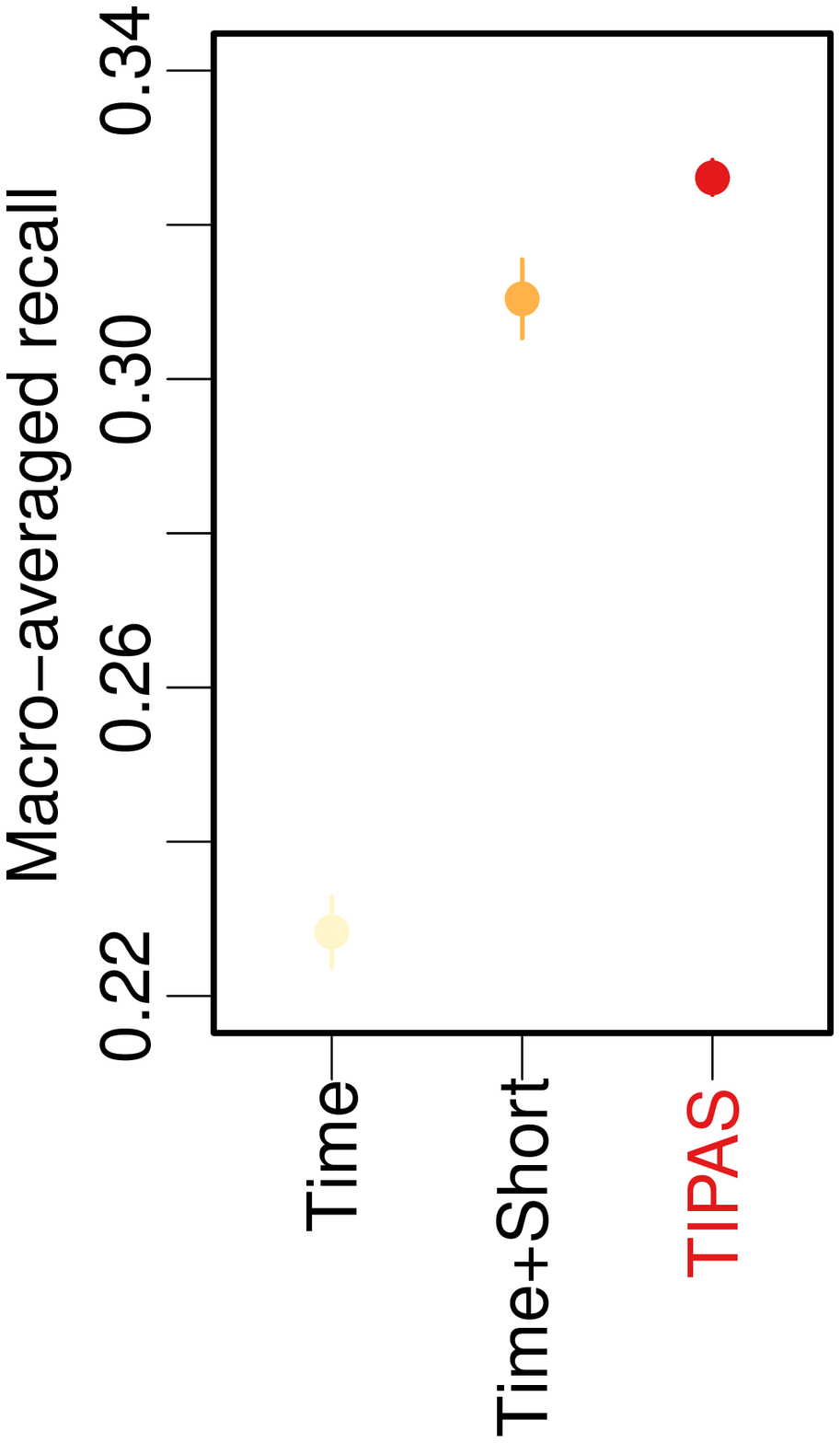}}
  \vspace{-3mm}
  \caption{
  Macro-averaged recall when predicting action. Higher is better.
  Comparing individual model components.
  }
  \label{fig:macro_avg_recall_components}
\end{figure}

We further compared our TIPAS model to several baselines on the macro-averaged recall metric (MAR) when predicting the next action.
This metric highlights differences in predictive performance on rare actions that do not affect the standard accuracy measure very much.
For example, more rare actions in both Argus and Under Armour dataset include walking, running, and biking (\eg, walking, running, and biking actions make up 0.05, 0.04, and 0.01 of the Argus dataset, respectively).
Variation in MAR highlights how well different models predict these rare actions.

\xhdr{Comparison to baseline models.}
As shown in Figure~\ref{fig:macro_avg_recall_baselines}, the eleven baseline models achieve MAR of 10.0-31.3\% on the Argus dataset and 12.5-29.9\% on the Under Armour dataset with different baselines performing best in each of the datasets (RNN and HMM, respectively).
Our full model TIPAS (Time+Short +Long) outperforms all baselines on both Argus (35.6\%; 14-256\% rel. improvement) and Under Armour datasets (32.6\%; 9-161\% rel. improvement).
These relative improvements in MAR over baselines are larger than those for accuracy (6-69\% and 11-156\% for Argus and Under Armour datasets, respectively).
This demonstrates that TIPAS can better support various types of user actions that are more rare compared with other baseline models.

\xhdr{Comparison to individual model components}
Note that our proposed model has three components (Equation~\ref{eq:lambda}): 
time-varying action propensities (Time), short-term interdependencies between actions (Short), and long-term periodic effects (Long).
Here, we evaluate the performance of each of these components in an ablation study by comparing Time, Time+Short, and the full proposed model TIPAS (Time+Short+Long) in terms of macro-averaged recall (MAR; Figure~\ref{fig:macro_avg_recall_components}; all models include user personalized preferences $\alpha_{ua}$). 
We find that modeling time-varying action propensities achieves a MAR of 27.2\% and 22.8\% on the two datasets, respectively.
Further, modeling short-term dependencies between actions improves this to 33.2\% and 31.0\%,
and capturing long-term periodicities of actions further improves this to 35.6\% and 32.6\%, respectively.
This demonstrates that capturing all three properties is essential to predicting actions in both datasets of human real-world action sequences.
Note that this is a bigger difference between the full Time+Short+Long model and the Time+Short model in terms of macro-averaged recall compared to instead using the accuracy measure ({7\% and 5\%} relative MAR improvements vs {3\% and 4\%} relative improvements in accuracy on the Argus and Under Armour datasets, respectively).
This indicates that modeling long-term periodicities is especially important for more rare actions.

\end{document}